\documentclass[11pt]{article}
\usepackage[utf8]{inputenc}
\usepackage{caption}
\usepackage{subcaption}
\usepackage[dvipsnames]{xcolor}
\usepackage{hyperref}
\usepackage[T1]{fontenc}
\usepackage{amsmath}
\usepackage{mathtools}
\usepackage{tensor}
\usepackage{braket}
 \usepackage{amsfonts}
\usepackage{setspace} 
\usepackage{amsthm}
\usepackage{graphicx}
\usepackage{xcolor}
\usepackage[section]{placeins}
\usepackage[normalem]{ulem}
\usepackage[margin = 2.5cm]{geometry}
\usepackage{amssymb}
\usepackage{comment}

\def\CI{{\cal I}}
\def\nref#1{(\ref{#1})}

\theoremstyle{definition}

\def\be{\begin{equation}}
\def\ee{\end{equation}}
\def\la#1{\label{#1}}
\def\bea{\begin{eqnarray}}
\def\eea{\end{eqnarray}}

\begin{document}

\thispagestyle{empty}
\begin{center}
    ~\vspace{5mm}

  \vskip 2cm 
  
   {\LARGE \bf 
      Logarithmic gravity from a very slow roll \\ \vspace{4pt} limit of inflation  
   }

    \vspace{0.5in}

 Jordan Cotler$^1$, Victor Ivo$^2$, and Juan Maldacena$^3$
  
    \vspace{0.5in}
  $^1$
{\it Department of Physics, Harvard University, Cambridge, MA 02138, USA}
   \\
   ~
   \\
  $^2$
{\it  Jadwin Hall, Princeton University,  Princeton, NJ 08540, USA }
   \\
   ~
   \\
  $^3$
{\it   Institute for Advanced Study,  Princeton, NJ 08540, USA }

\end{center}

\vspace{0.5in}

\begin{abstract}
We study the inflationary no-boundary wavefunction in a ``linear-roll limit,'' in which $G_N \to 0$ while the Hubble scale $H$ and the slope of the inflaton potential are held fixed. Although bulk graviton fluctuations are suppressed in this limit, quantum inflaton fluctuations remain finite. Because reheating occurs on a surface of fixed inflaton value, these fluctuations become fluctuations of the conformal factor of the reheating metric. The probability measure for these fluctuations simplifies drastically and becomes Gaussian, even for large fluctuations. The resulting log-correlated theory at superhorizon scales is a $d$-dimensional analog of Liouville gravity without the exponential potential, which we call Logarithmic Gravity. The quadratic and linear terms are determined by conformally covariant scattering data; in even dimensions we relate them to the critical GJMS operator and Branson $Q$-curvature. From bulk unitarity, we argue that one-loop effects do not generate additional conformal-factor dependence in the physical probability measure, and we verify the cancellation explicitly in $2 + 1$ dimensions. For the no-boundary norm with a rolling inflaton, we show that the noncompact residual conformal symmetries can be gauge fixed with a finite Faddeev-Popov determinant, so they do not force the sphere contribution to vanish. In addition, conditioning on the absence of vacuum decay by bubble nucleation generates the Liouville exponential potential in the probability measure.
\end{abstract}
 
\vspace{1in}

\pagebreak

\setcounter{tocdepth}{3}
{\hypersetup{linkcolor=black}\tableofcontents}

\section{Introduction}
\la{Intro}
 
Inflation provides a probability distribution for the geometry of the reheating surface. We consider a limit in which this random-geometry problem simplifies dramatically. Namely, we take $G_N \to 0$ while keeping the slope of the inflaton potential constant \cite{Creminelli:2008es,Chen:2024rpx}. This freezes the tensor fluctuations but allows scalar fluctuations of arbitrary size. We show that the no-boundary probability functional for the conformal factor of the metric on the $d$-dimensional reheating surface, $g_{ij} = e^{ 2 \zeta } \hat \gamma_{ij} $, at superhorizon scales becomes\footnote{Please, do not confuse $Q$ with ${\cal Q}_{\hat \gamma}$. (This is an unfortunate, historically rooted clash of notation.)} 
\be \la{LogGra}
p[\zeta , \hat \gamma ] = 
\exp \left\{ -C_d Q^2  \int d^d x \sqrt{ \hat \gamma } \left[ \zeta {\cal P}_{\hat \gamma } \zeta + 2 {\cal Q }_{\hat \gamma } \zeta \right]  \right\} ~,~~~~~~ Q^2= { \dot \phi^2  \over H^{d+1} } {\rm Vol}(S^{d+1}) 
\ee 
where $C_d$ is a numerical constant.
${\cal P}$ is an operator that goes as ${\cal P} \approx |k|^d$ at short distances and ${\cal Q} $ is a kind of curvature of the manifold. Although the inflaton becomes free in this limit \cite{Creminelli:2008es}, using it as the clock that defines the reheating surface converts its fluctuations into fluctuations of the conformal factor of the reheating metric. The nontrivial point is that the resulting probability functional remains Gaussian even for arbitrarily large $\zeta$. Thus, in this limit, the $d+1$-dimensional problem of determining the scalar geometry of the reheating surface reduces to the simpler $d$-dimensional scalar gravity theory \nref{LogGra}.    

This action is invariant under conformal transformations treating $\zeta$ as a logarithmic dimension zero field \cite{Gurarie1993, Zamolodchikov2004}.  
It is a $d$ dimensional generalization of the familiar two dimensional Liouville action without the cosmological constant term\footnote{In two dimensions this is sometimes called ``linear dilaton''.}, very similar (identical in many cases) to the ones discussed previously in the physics \cite{Levy:2018bdc, Kislev:2022} and mathematics literature \cite{Cercle:2019jxx, Schiavo:2021unx}.  What is notable in our setting is that this random geometry theory is not introduced as an independent $d$ dimensional model, but rather arises directly as the probability measure on the reheating surface of an ordinary $d+1$ dimensional inflationary theory.  In particular, this gives a physical realization of the volume measures studied in the ``Gaussian multiplicative chaos'' literature \cite{Rhodes:2013iua}. $\zeta$ has a  logarithmic two point function 
at short distances, which leads to interesting features for composite operators such as the volume, $e^{d \zeta}$ \cite{Creminelli:2008es, Dubovsky:2008rf, Cotler:2026lna}. 

The operator ${\cal P}_{\hat \gamma} $ appearing in \nref{LogGra} is generically non-local. However, for the de Sitter filling and $d$ even, it becomes local and equals the GJMS operator \cite{Graham:1992gjms}. Thus, in this case, the structure of the probability measure is directly related to familiar objects in conformal geometry.  This no-boundary prescription also gives us the linear term, which has a well known phenomenological problem: it pushes the volume of the reheating surfaces towards small values. We will not address this familiar problem. Note that other choices of the wavefunction can avoid this problem and are expected to yield the same quadratic term at short scales, though this does not address the problem within the no-boundary prescription itself.

It is also possible to modify the physical setup so as to obtain the missing Liouville potential. We consider an inflationary vacuum that can decay by bubble nucleation \cite{Coleman:1980aw} and condition on the event that it has {\it not} decayed by the time of reheating. The resulting no-decay probability generates the Liouville exponential potential.

We also consider one-loop effects. These are potentially important in the linear-roll limit because the tree-level inflaton action remains of order one as $M_{\rm pl}\to\infty$, so one-loop effects can contribute at the same order to the probability measure for $\zeta$. We argue, however, that they do not generate any additional conformal-factor dependence in the physical probability measure. We demonstrate this explicitly in $2+1$ dimensions, where it follows from an interesting interplay between measure effects and one-loop corrections, ultimately reflecting bulk unitarity.

 The origin of this cancellation can be understood by considering the cutoff.  Short-distance modes are regulated at a fixed physical momentum scale. Expressed instead relative to the fiducial, or comoving, metric, the corresponding cutoff depends on $\zeta$; the resulting $\zeta$ dependence of the one-loop factors is canceled by that of the inner product measure. Similar results are expected in higher dimensions, although they have not been checked explicitly.

The one loop analysis also involves a gauge fixing of conformal reparametrizations. In the pure gravity case around the de Sitter solution, division by the volume of this non-compact group sets the norm to zero \cite{Cotler:2025gui}. In our case, the reparametrizations act nontrivially on $\zeta$ and can be used to remove the $l=1$ modes with a finite Faddeev-Popov determinant; they do not set the norm to zero. In other words, inflation makes the norm non-zero.

In our limit,  on the reheating surface, we have a theory of gravity where the only dynamical field is the scale factor. As an amusing historical note,  this is similar to Nordstr\"om gravity
\cite{Einstein:1914bu}, except that here the kinetic term is different. There are several papers that consider cosmology and inflation based on Liouville gravity in the {\it bulk}, see e.g. \cite{CarneirodaCunha:2003mxy,Anninos:2024iwf}.  

The paper is organized as follows. In section \ref{sec:NB} we describe the setup in more detail and derive \nref{LogGra}, defining all the operators carefully. In section \ref{loops},  we discuss the effects of one loop corrections. The main conclusion is that they have  no net  effects on  the $\zeta$ dependence in \nref{LogGra}. In section \ref{sec:Scat}, we discuss in more detail ${\cal P}_{\hat \gamma}$ and ${\cal Q}_{\hat \gamma}$ as well as the connections with previous mathematical literature which defines similar operators in the anti-de Sitter context \cite{Graham:2003GZ, Fefferman:2002QCurvaturePoincare}. In section \ref{sec:GA}, we mention that the curvature ${\cal Q}_{\hat \gamma} $ also appears in the evaluation of the pure gravity action, as was discussed in the AdS case \cite{Fefferman:2002QCurvaturePoincare}. We also make a few comments about KSW \cite{Kontsevich:2021dmb, Witten:2021nzp} and explain how to add a Liouville potential to \nref{LogGra}.

\section{No-boundary wavefunctional of linear roll inflation}
\la{sec:NB}
\subsection{Generalities about inflation and no-boundary geometries}
\label{gennb}

\begin{figure}[t!]
    \centering
    \includegraphics[width=0.75\linewidth]{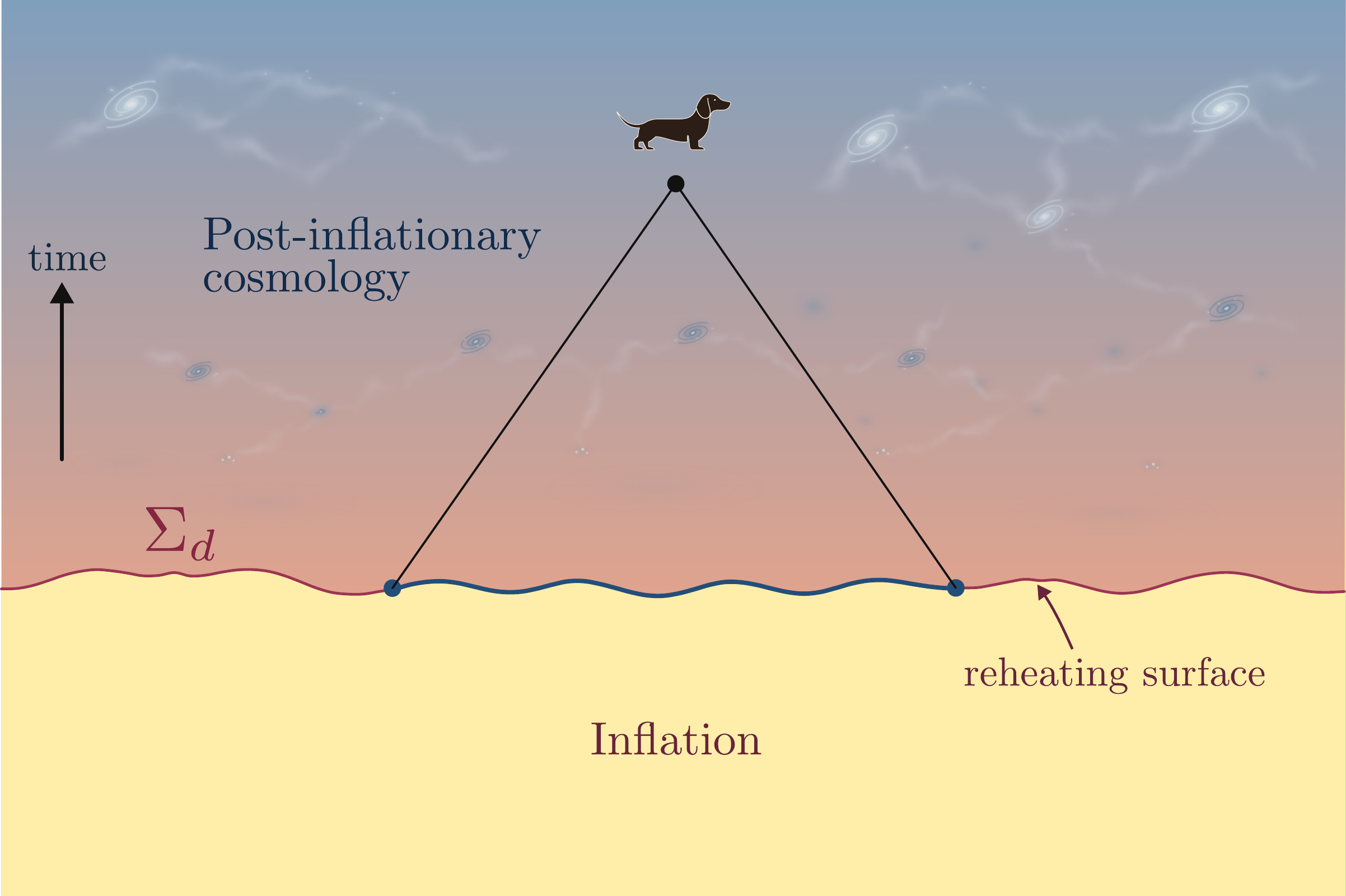}
    \caption{Cartoon of the Penrose diagram for $d+1$ dimensional cosmologies studied in inflation. There is a period of inflation at very early times, which ends at a reheating surface $\Sigma_d$. The surface $\Sigma_d$  is observable (indirectly) to someone (the dachshund) living in the later universe. One role of inflation is to provide a probability measure for the random geometry of $\Sigma_d$.}
    \label{cosmogic}
\end{figure}

We consider a cosmology with an early period of cosmological inflation; see Figure \ref{cosmogic}. We include quantum effects during inflation, but we will treat the post inflationary cosmology classically. We view this later period as a measurement apparatus that is measuring the inflationary wavefunctional, which is the main object we will study. 

During the inflationary period, the effective description consists of gravity minimally coupled to an inflaton $\phi$, with the Lorentzian action 
\begin{equation}
\label{Ilor}
I_{L}=\frac{M_{\rm pl}^{d-1}}{2}\int_{\mathcal{M}} d^{d+1}x\sqrt{-g}\, R+\int_{\mathcal{M}} d^{d+1}x\,\sqrt{-g}\bigg[-\frac{1}{2}(\nabla \phi)^{2}-V(\phi)\bigg]+\text{(boundary terms)}
\end{equation}
with $\mathcal{M}$ the bulk spacetime, and we leave the discussion of boundary terms for later. The scalar field rolls down the potential while also experiencing quantum fluctuations. When the field reaches a special value $\phi = \phi_r$, inflation ends. For conceptual simplicity, we assume that the end of inflation is sudden, though it could extend a few Hubble times without major change to our discussion. The end of inflation occurs on a spacelike surface $\Sigma_d$, called the ``reheating surface''. This surface has some fluctuations which originate as quantum fluctuations during the inflationary evolution. We will be making statements about the probability distribution for the shape of this surface. 

\begin{figure}[t!]
    \centering
    \includegraphics[width=0.7\linewidth]{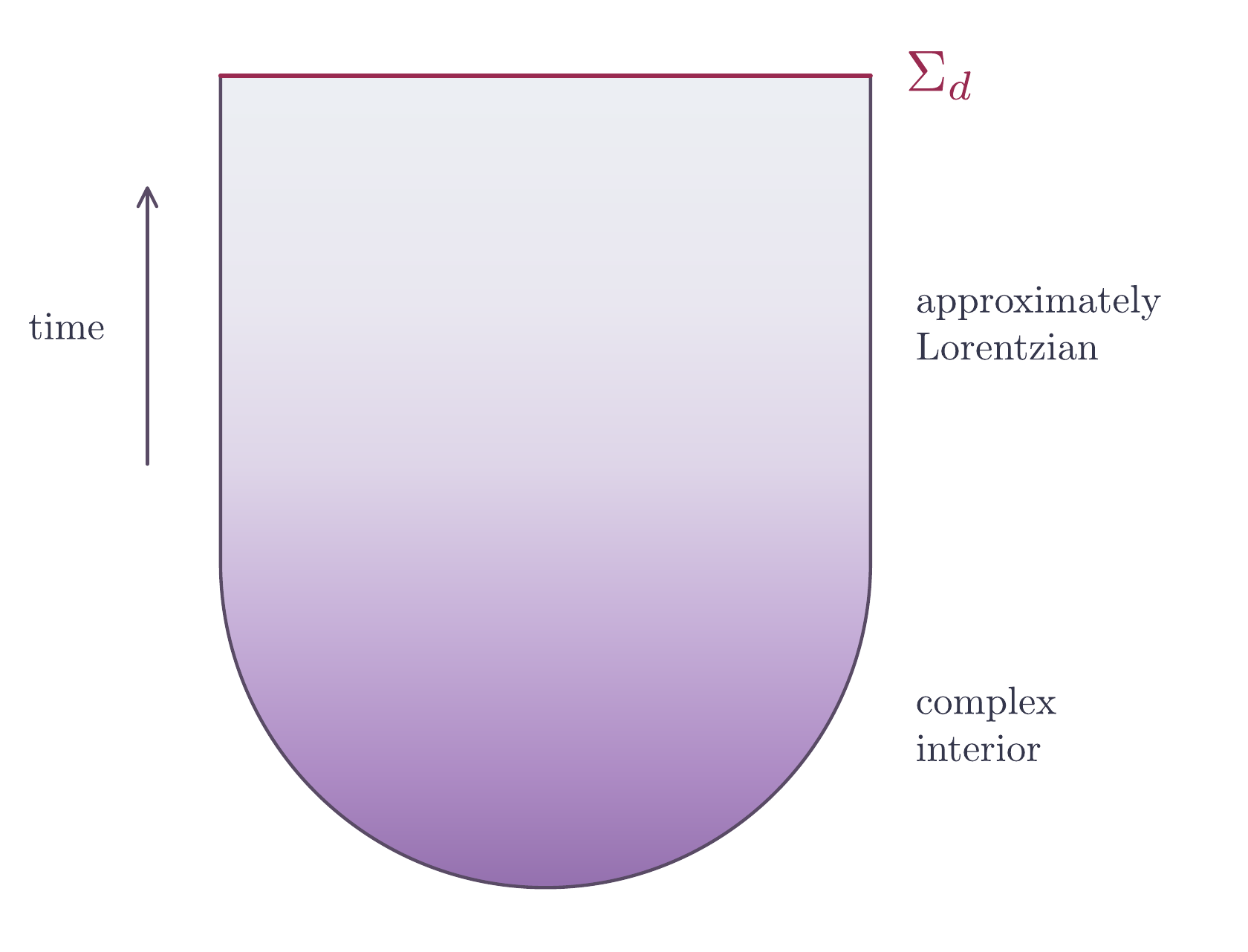}
    \caption{An example of a no-boundary geometry we consider in this paper. The color indicates how ``complex'' the geometry is. Near the boundary, and generally at late times, the geometry is almost Lorentzian with small complex corrections. As one goes back in time, the geometry becomes genuinely complex, and eventually caps off.}
    \label{hhfig}
\end{figure}

There are various methods to compute this probability and they all tend to agree for the probability of short wavelength fluctuations. Here we will compute it using the no-boundary proposal \cite{Hartle:1983ai}, which has certain phenomenological successes and failures; see 
\cite{Maldacena:2024uhs} for a short review.  The proposal involves the following procedure. Given a particular $d$ dimensional spatial geometry $\Sigma_d$ and scalar field profile $\phi_{b}(x)$, we integrate over generically complex bulk $d+1$ dimensional spacetimes with a boundary at the given geometry and no further boundary, see Figure \ref{hhfig}, 
\begin{equation}
\label{nbdry}
\Psi[\Sigma_{d},\phi_{b}]=\int Dg \,D\phi\,e^{i I_{L}}
\end{equation}
As is standard in semiclassical gravity, this is performed by first finding a classical geometry and then integrating over small fluctuations around it. As we will discuss,   for our problem the important contribution comes from the classical geometry. The bulk manifolds are close to the usual Lorentzian solutions near $\Sigma_d$ but have small complex deformations that grow to larger values for early times. These are necessary in order to have a bulk solution with no extra boundary, and no singularity.  

The focus of our work is on a further restriction to a ``linear-roll'' limit of inflation, introduced in \cite{Creminelli:2008es,Chen:2024rpx},  
\begin{equation} \la{LinRol}
\textbf{Linear Roll limit}:~~M_{\rm pl}\rightarrow \infty~\text{ and }~H,~\partial_{\phi}V = \text{constant}
\end{equation}
For simplicity we will take $H=1$ from now on, restoring it whenever necessary. The limit \nref{LinRol} can also be viewed as a very slow roll limit, where the slow roll parameters are much smaller than what is strictly necessary for ordinary slow roll inflation. This limit can be achieved by taking the following inflaton potential 
\begin{equation} \la{LinPot}
V(\phi)=V_0 + (\phi-\phi_r)V' ~,~~~~{\rm with }
~~~ V_0 \equiv \frac{d(d-1)}{2}M_{\rm pl}^{d-1} 
\end{equation}
with $V'$ a fixed number as $M_{\text{pl}} \rightarrow \infty$. Due to the form of the potential in \nref{LinPot}, the action of $\phi$ is just that of a free massless scalar with a linear source term, coming from $V'$. The $V_{0}$ term survives as a positive cosmological constant. Due to the $M_{\rm pl} \rightarrow \infty$ limit, the gravitational path integral in \nref{nbdry} localizes around classical bulk geometries  which  are Einstein manifolds   
\begin{equation}
\label{eineq}
R_{\mu \nu}=d \, g_{\mu\nu} ~,~~~~~~~~{\rm for } ~~~H=1
\end{equation}
The classical equation of motion for the scalar $\phi$ is 
\begin{equation}
-\nabla^{2}\phi+V'=0
\end{equation}
This scalar field can have large quantum fluctuations but, due to \nref{LinRol}, these do not backreact on the bulk geometry. In other words, the bulk gravity solution in \nref{eineq} provides a fixed classical background where $\phi$ propagates. The no-boundary wavefunctional in \nref{nbdry} therefore has the form
\begin{equation}
\label{psiket}
\Psi[\Sigma_{d},\phi_b]\approx \sum_{\text{saddles}} A_{+}e^{i(I_{\text{grav,+}}+I_{\phi,+})}
\end{equation}
with $A_{+}$ the product of one-loop corrections for the gravity degrees of freedom and the scalar ones. $I_{\text{grav},+}$ and $I_{\phi,+}$ are the on-shell gravity and scalar actions respectively, the $+$ stands for the fact that we are computing $\Psi$, which is a ``ket''. There are similar solutions that compute the ``bra''  $\Psi^{*}$,   which we denote with a minus index. The ket manifold involves a complex deformation at early times in the positive imaginary time direction, whereas in the bra manifold, the complex deformations are in the negative imaginary time direction; see section \ref{sphex} for a particular example, and \cite{Witten:2021nzp} for further discussion.  

The explicit form of $I_{\text{grav},+}$ is given by
\begin{equation} \la{GravAct}
I_{\text{grav},+}=M_{\rm pl}^{d-1}\bigg[d\int_{\mathcal{M}} d^{d+1}x\sqrt{-g}-\int_{\Sigma}d^{d}x\,\sqrt{g_b}\,K \bigg]
\end{equation}
with $g_b$ the metric of the boundary surface $\Sigma_d$ and $K$ its extrinsic curvature, where we used the Einstein equation to simplify the bulk term.  The action for $I_{\phi,+}$ is given by\footnote{One might worry about the following. The backreaction in the metric due to the $\phi$ solution, which is of order $\delta g \propto 1/M_{\rm pl}^{d-1}$, could give an order one term when inserted in the gravity part of the action. Fortunately, since the original metric is a classical solution, this first order change vanishes.}
\begin{equation}
\label{Iphi+}
I_{\phi,+}=\int_{\mathcal{M}}d^{d+1}x\,\sqrt{-g} \bigg[-\frac{1}{2}(\nabla \phi)^{2}-(\phi-\phi_{r})V'\bigg]
\end{equation}
   Note that we first solve the pure gravity problem with a fixed cosmological constant, using \nref{eineq} and compute its action \nref{GravAct}. This part of the problem is exactly the same as that of a pure gravity problem with a fixed cosmological constant. Then we solve the problem of a scalar field on a fixed background geometry and evaluate its action \nref{Iphi+}, but it is important that we use $\phi$ as a clock to determine the position of the surface.

We are interested in studying $|\Psi|^{2}$. This will include a double sum over saddles, the ``bra'' and ``ket'' saddles. We can therefore write $|\Psi|^{2}$ as  
\begin{equation}
\label{psibraket}
|\Psi|^{2}=\sum_{\text{bra},\text{ket}} A_{+}A_{-}e^{-\CI }
\end{equation}
where the sum is over bra and ket saddles, and $\CI $ is the classical bra-ket weight of a given pair. It contains both a gravitational and a scalar piece, $\CI_{\text{grav}}$ and $\CI_{\phi}$, each of which has the structure
\begin{equation}
\label{Wdef}
\CI =-i(I_{+}-I_{-})
\end{equation}
with $I_{-}$ the respective action of the bra saddle. 

Let us discuss how the terms $\CI_{\text{grav}}$ and $\CI_{\phi}$ depend on the boundary conditions. We are interested in evaluating the wavefunction of the universe at a large surface $\Sigma_d$. For Einstein spacetimes \nref{eineq}, this implies that the geometry extends well into its asymptotic region, where it takes the form \cite{Starobinsky:1982mr, Skenderis:2002wp}
\begin{equation}
\label{metfg}
ds^{2}=\frac{-d\eta^{2}+\gamma_{ij}(\eta) dx^i dx^j}{\eta^{2}}
\end{equation}
with $\gamma_{ij}(\eta)$ approximately real for small $\eta$, which corresponds to late times. The surface $\Sigma_d$ is then  at small $\eta_{b}$, where the boundary value of the field $\phi_{b}$ and the surface metric $g_{b}$ are  
\begin{equation}
\label{metbc}
g_{b\, ij}=\frac{\gamma_{b\, ij}}{(-\eta_{b})^{2}}=\frac{\gamma_{ij}(\eta_{b})}{(-\eta_{b})^{2}}~~,~~\phi_{b}=\phi(\eta_{b})
\end{equation}
where $\gamma_{b\, ij}=\gamma_{ij}(\eta_{b})$ fixes the boundary metric, since we fixed $\eta_{b}$. It is common then to think of the ``renormalized metric'' $\gamma_{b\, ij}$ as the quantity we fix. Of course, the physical metric is $g_{b \, ij}$. 

The bulk theory is invariant under local time reparametrizations, $t \to t + \omega(x)$. At late times, these  reparametrizations change 
\begin{equation}
\label{weyl1}
\gamma_{b \, ij} \rightarrow e^{2\omega}\gamma_{b\, ij}~~,~~~~~~~~\phi_{b}\rightarrow\phi_b + \dot \phi \omega=\phi_{b}-\frac{V'}{d}\omega  
\end{equation}
Since this is a bulk gauge symmetry asymptotically, it should not change the probabilities. Indeed, this is the case when we use the no-boundary prescription (see \cite{Hartle:2008ng} and \cite{Ivo:2024ill} for more details).  Mathematically, this follows from the Hamiltonian constraint when the universes are large. Furthermore, the two configurations in \nref{weyl1} lie along a common real classical solution at late times. 

In fact, since the wavefunctional for the gravity and scalar degrees of freedom factorizes, and each independently defines a well-defined wavefunctional problem, then $\CI_{\text{grav}}$ and $\CI_{\phi}$ should be independently invariant under \nref{weyl1}. However, $\CI_{\text{grav}}$ does not depend on $\phi$, so the classical probability weight for the gravity degrees of freedom should be invariant under Weyl rescalings of $\gamma_{b}$. This is just the statement that the degrees of freedom of pure gravity correspond to the metric of $\Sigma_d$ up to Diffeomorphism and Weyl transformations \cite{Chakraborty:2023yed}. This is why the physical fluctuations of $\Sigma_d$ in pure gravity are those of gravitons.  On the other hand, $\CI_{\phi}$ can in principle depend on the metric and $\phi$. 
 
When we evaluate probabilities, we should therefore choose a representative along this gauge orbit. We will do this by fixing the inflaton on the boundary to its reheating value, $\phi_b = \phi_r$. The remaining scalar degree of freedom is then encoded in the local scale factor of the reheating metric
\begin{equation}
\label{bdrycd}
\gamma_{b\, ij}=e^{2\zeta}\hat{\gamma}_{b \, ij}~~,~~~~~~~~~~~~\phi|_{\Sigma_d}=\phi_{r}
\end{equation}
where $\zeta$ corresponds to the ``scalar fluctuations'', and $\hat{\gamma}_{b\, ij}$ is a metric specified in the conformal class of $\Sigma_d$. The action for $\zeta$ comes from $\CI_{\phi}$ alone, so that $\zeta$ can have large fluctuations. However, the action for tensor fluctuations of $\hat{\gamma}_{b \, ij}$  is  dominated by $\CI_{\text{grav}}$, and, as $M_{\text{pl}}\rightarrow \infty$, fluctuations away from the global minimum of $\CI_{\text{grav}}$ will be heavily suppressed. This is the statement that graviton fluctuations go to zero, and the conformal class $\hat{\gamma}_{b}$ of the reheating surface is ``pinned'' to the absolute minimum of $\CI_{\text{grav}}$. 

For most purposes, we will assume that $\hat{\gamma}_{b \, ij}$ is fixed, and we are studying fluctuations of $\zeta$. This is well defined if $\hat{\gamma}_{b\, ij}$ is at its global minimum, or if we imagine that we leave the integral over $\hat{\gamma}_{b\, ij}$ last. For some of the cases we consider, the dominant saddle is de Sitter space, but we discuss other possibilities in section \ref{PureGrav}.  

Note that the metric decomposition \nref{bdrycd} has a trivial gauge symmetry 
\be \la{Gsym}
\zeta \to \zeta -  \sigma ~,~~~~~~~ \hat \gamma_{ij} \to e^{ 2 \sigma } \hat \gamma_{ij}
\ee 
where $\sigma = \sigma(x)$ is a general function on the manifold. This transformation leaves the physical boundary metric $\gamma_{b\,ij}=e^{2\zeta}\hat\gamma_{b\,ij}$ unchanged; it is simply an ambiguity in how we split the metric into a scale factor and a fiducial metric. We will therefore require all our expressions to be invariant under \nref{Gsym}. We sometimes call $\hat\gamma_{ij}$ the ``fiducial'' metric.

This redundancy is distinct from the asymptotic bulk gauge symmetry \nref{weyl1}. The latter changes the boundary data $(\gamma_b,\phi_b)$ along a bulk gauge orbit and has already been fixed by imposing $\phi_b=\phi_r$. By contrast, \nref{Gsym} acts only on the decomposition of the fixed reheating metric $\gamma_b$ into $\zeta$ and $\hat\gamma_b$. 

A side comment is that the situation is somewhat reminiscent to that encountered in the discussion of near extremal black holes or nearly-AdS$_2$ gravity, or JT gravity \cite{Almheiri:2014cka, Maldacena:2016upp, Engelsoy:2016xyb, Jensen:2016pah, Ferrari:2024kpz} . In that case, there is a fixed metric, which is AdS$_2$, and a scalar field whose value sets the physical position of the boundary. One difference is that here the scalar is a propagating field in the bulk.

\subsection{No-boundary wavefunction for the inflaton}
\label{nbdinfl}

Here we compute the classical action $\CI_{\phi}$  in section \ref{gennb} as a function of $\zeta$ on the reheating surface, for a given value of $\hat{\gamma}_{b}$ in \nref{bdrycd}, in the regime that the surface $\Sigma_{d}$ is large. While the computation of small scalar fluctuations is completely standard, here we give a derivation which is valid for arbitrarily large $\zeta$, something possible thanks to the limit \nref{LinRol}.  Since $\CI_{\text{grav}}$ is independent of $\zeta$ it is enough to compute $\CI_\phi$.  
 
The invariance of $\CI_\phi$ under the asymptotic time reparameterization \nref{weyl1} relates its dependence on the scale factor to its dependence on the boundary value of the inflaton. Since an infinitesimal transformation acts as $\delta\zeta = \omega$ and $\delta\phi_b = - (V'/d)\omega$, we expect that
\be \la{ZetaWder}
{ \delta \CI_\phi \over \delta \zeta(x) } =  {    V' \over d } { \delta \CI_\phi \over \delta \phi_b(x) }
\ee 
Since this is an important equation, we check it explicitly in appendix \ref{CheckTime}.  
We then use that the $\phi_b$ derivative on a classical solution gives 
\be \la{PhiWder}
{ \delta \CI_\phi \over \delta \phi_b(x) } = - i { \sqrt{\gamma_{b}} \over (-\eta_b)^{d} } ( \partial_n \phi_{+} - \partial_n \phi_-) 
\ee 
In order to compute these derivatives, it is useful to define an auxiliary variable $N$ via 
\begin{equation}
\label{Ndef}
N=-\frac{d(\phi_{r}-\phi)}{V'}
\end{equation}
with the equation of motion and boundary conditions
\begin{equation}
\label{Neq}
\nabla^{2}N=d~~,~~ N|_{\Sigma}=0
\end{equation}

At large times, when $\phi$ approaches the slow-roll trajectory, the variable $N$ decreases at a rate of one per e-fold. Therefore, since $N=0$ defines reheating, $N$ is a variable that measures how many classical e-folds $\phi$ would locally  remain until reheating. Equation \nref{Neq} implies that $N$ goes as $\log(-\eta)-\log(-\eta_{b})$ at small $\eta$. Moreover, the differential equation in \nref{Neq} defines perturbative small-$\eta$ corrections to $N$, which for low enough order are local functions of $g_{b}$. Corrections of order $\eta^{d}$, however, are filling dependent and determined by the regularity of $N$ in the interior of the geometry. See section \ref{eqwvfct} for more details. This implies that the normal derivatives of $N$ at $\Sigma$ behave as\footnote{$ 
\operatorname{Vol}(S^d) = 2\pi^{\frac{d+1}{2}}
/\Gamma\!\left(\frac{d+1}{2}\right) 
 $.} 
 \bea
\partial_{n}N_{+}+\partial_{n}N_{-}&=&-2+ \cdots  \la{NormDerNFl} \\ 
\partial_{n}N_{+}-\partial_{n}N_{-}&=&\frac{2i dB_{d}}{\Gamma(d)}(-\eta_{b})^{d}\mathcal{Q}_{\gamma_{b}}+\cdots  ~,~~~~{\rm with }~~~B_{d}\equiv \frac{\text{Vol}(S^{d+1})}{2\text{Vol}(S^{d})} \la{norm+-}
\eea
where the dots stand for terms subleading at small $\eta_{b}$, and $\mathcal{Q}_{\gamma_{b}}$ is a response function that measures the difference between filling dependent terms in the bra and the ket. 
For our current purposes, $\mathcal{Q}_{\gamma_{b}}$ is simply defined from equation \nref{norm+-}. Note that it depends non-trivially on the metric on $\Sigma_d$, including its scale factor $\zeta$. It also depends on the bulk filling of the bra and ket, but we omit that dependence to avoid clutter.  The numerical prefactors in \nref{norm+-} were picked for later convenience. 

Substituting \nref{norm+-} into \nref{PhiWder} and \nref{ZetaWder} we obtain 
\begin{equation}
\label{diffW}
\frac{\delta \CI_{\phi}[\zeta, \phi_r]}{\delta \zeta}=\frac{2dB_{d}}{\Gamma(d)}\bigg(\frac{V'}{d}\bigg)^{2}\sqrt{\gamma_{b}}\mathcal{Q}_{\gamma_{b}} 
\end{equation}
to leading order in small $\eta_b$. 

Conceptually, $\mathcal{Q}_{\gamma_b}$ measures the difference between the bra and ket responses in the sourced problem \nref{Neq}. It is now useful to state how $\mathcal{Q}_{\gamma_{b}}$ transforms as we Weyl-rescale $\gamma_{b}=e^{2\zeta}\hat{\gamma}_{b}$. For now, we just quote the result and discuss it in more detail in section \ref{folweyl}.
For this purpose, it is necessary to introduce an associated linear operator $\mathcal{P}_{\gamma_{b}}$. The operator $\mathcal{P}_{\gamma_{b}}$ is defined from a problem similar to \nref{Neq}, but which is instead defined for a function $f$ via the Dirichlet problem
\begin{equation}
\label{feq}
\nabla^{2}f=0~~,~~f|_{\Sigma}=f_{b}
\end{equation}

For the same reasons as in \nref{Neq}, the first filling dependent term of $f$ in the small $\eta$ expansion is of order $(-\eta)^{d}$. This term must be linear in $f_{b}$ since the problem in \nref{feq} is linear. Therefore, for a pair of solutions $f_{\pm}$ for a ket and a bra, we can define
\begin{equation}
\label{fdisc}
\partial_{n}f_{+}-\partial_{n}f_{-}=\frac{2i dB_{d}}{\Gamma(d)}(-\eta_{b})^{d}\mathcal{P}_{\gamma_{b}}f_{b} + \cdots 
\end{equation}
where $\mathcal{P}_{\gamma_{b}}$ is a linear operator, and $B_d$ the same as in \nref{norm+-}. Note that, like $\mathcal{Q}_{\gamma_{b}}$, the operator $\mathcal{P}_{\gamma_{b}}$ depends on the bulk filling of the bra and ket, but we omit that dependence here. We can then show (see section \ref{folweyl}) that $\mathcal{Q}_{\gamma_{b}}$ and $\mathcal{P}_{\gamma_{b}}$ change under Weyl transformation as
\begin{equation} \la{QPRel}
\mathcal{Q}_{e^{2\omega}\gamma_{b}}=e^{-d\omega}(\mathcal{Q}_{\gamma_{b}}+\mathcal{P}_{\gamma_{b}}\omega)~~,~~ \mathcal{P}_{e^{2\omega}\gamma_{b}}=e^{-d\omega}\mathcal{P}_{\gamma_{b}}
\end{equation}

Using this result, we can integrate \nref{diffW} to obtain
\bea
\la{WActL}
\CI_{\phi}[e^{2\zeta}\hat{\gamma}_{b}]&=&\CI_{\phi}[\hat{\gamma}_{b}]+{\cal I}_{\ell}[\zeta, \hat \gamma ] ~,~~~~{\rm where}
\\
  {\cal I}_{\ell}[\zeta, \hat \gamma ]&=&\frac{dQ^{2}}{2\Gamma(d)\text{Vol}(S^{d})}\int_{\Sigma}d^{d}x\,\sqrt{\hat{\gamma}_{b}}(\zeta \mathcal{P}_{\hat{\gamma}_{b}}\zeta+2 \mathcal{Q}_{\hat{\gamma}_{b}}\zeta)  \label{Wliouv} 
\eea
with 
\begin{equation}
Q^{2}=\bigg(\frac{V'}{d}\bigg)^{2}\text{Vol}(S^{d+1})= { \dot \phi^2 \over H^{d+1} }\text{Vol}(S^{d+1})
\end{equation}
defined as in \cite{Cotler:2026lna} (see equation (62) there) and we restored $H$ in the last formula. 

 Equations \nref{WActL} and \nref{Wliouv} are the main tree-level result of this section. The dependence on $\zeta$ is exactly quadratic plus linear, even when $\zeta$ itself is not small. All dependence on the particular bra-ket filling is encoded in the response data $\mathcal{P}_{\hat\gamma_b}$ and $\mathcal{Q}_{\hat\gamma_b}$, while $Q^2$ fixes the overall strength of the fluctuations. So the large-fluctuation problem has reduced to a Gaussian theory, despite the nonlinear relation between $\zeta$ and the reheating metric. 

 It is also useful to express $Q^{2}$ in terms of the more usual slow-roll parameter $\epsilon$ of inflation and the de Sitter entropy $S_{\text{dS}}$ 
\begin{equation}
\label{epsdef}
S_{\text{dS}}=d\bigg(\frac{M_{\rm pl}}{H}\bigg)^{d-1}\text{Vol}(S^{d+1})~,~~\epsilon=\frac{\dot{\phi}^{2}}{(d-1)M_{\rm pl}^{d-1}H^{2}}~~~\Longrightarrow~~~Q^{2}=\frac{(d-1)}{d}\,\epsilon\,S_{\text{dS}}
\end{equation}
so we see that the regime $Q^{2}\sim O(1)$ corresponds to $\epsilon \sim S_{\rm dS}^{-1}$, which is very small in the large $M_{\rm pl}$ limit.
 
Note that short distance modes are not sensitive to the bulk filling, so for them the operator  ${\cal P}_{\hat \gamma_b}$ is the same as the one that appears when we compute the wavefunctional of a massless field in de Sitter space \cite{BRANDENBERGER1984328}. At short distances,
it behaves as 
\begin{equation}
\label{Pflat}
\mathcal{P}_{\hat{\gamma}_{b}}\approx |k|^{d} \approx (-\hat{\nabla}^{2})^{\frac{d}{2}}
\end{equation}
with $\hat{\nabla}^{2}$ the laplacian of the manifold with metric $\hat{\gamma}_{b}$. For odd $d$, as in the case of our universe, $\mathcal{P}_{\hat{\gamma}_{b}}$ is  a non-local operator. For other relevant properties of $\mathcal{P}_{\gamma_{b}}$, see section \ref{asygen}. 

The action \nref{Wliouv} then implies that the short distance two-point function of $\zeta$ behaves  as 
\begin{equation}
\label{zetatwopt}
\langle \zeta(x)\zeta(y)\rangle=\frac{2}{dQ^{2}}\log \frac{1}{|x-y|}+ \cdots
\end{equation}
The dots are terms subleading at short distances.

At tree level, \nref{Wliouv} is valid for fluctuations of arbitrary size and is not a small-$\zeta$ expansion. In section \ref{loops}, we argue that after the one-loop prefactors are combined with the boundary integration measure, they generate no additional $\zeta$ dependence in the physical probability measure.

From the trivial gauge symmetry \nref{Gsym} applied to both sides of \nref{WActL}, we expect 
\be 
{\cal I}_\ell[ \zeta - \sigma , e^{ 2 \sigma } \hat \gamma ]= {\cal I}_\ell[ \zeta   ,  \hat \gamma ]-{\cal I}_\ell[  \sigma ,   \hat \gamma ]
\ee 
This can be explicitly checked using \nref{QPRel}. As expected, this is precisely the transformation needed for \nref{WActL} to be independent of the choice of decomposition $\gamma_b = e^{2\zeta}\hat\gamma$. The transformation law has the form of a classical Wess-Zumino, or cocycle, relation associated with changing the fiducial representative $\hat\gamma$. In even dimensions, it is closely analogous to the transformation law of a conformal anomaly. This should be distinguished from the regulator-dependent one-loop Weyl anomaly discussed in section \ref{loops}. Notice that this is intimately related to the fact that the spectrum of scalar fluctuations is precisely scale invariant in our limit \nref{LinRol}.  Note that the symmetry \nref{Gsym} is a general symmetry even away from this particular limit, but it acts on the wavefunction in a more complicated way. In particular, it leads to non-linearities that are implied by the soft limits and ``inflationary consistency conditions'' \cite{Maldacena:2002vr, Creminelli:2004yq}. We emphasize that such non-linearities are proportional to the deviation from exact scale invariance, so they vanish in our case. See appendix \ref{MassFie} for further discussion. 

We should also mention that while we discussed in this section the wavefunctional for $\zeta$, a late time observer such as us only has access to part of the reheating surface; see Figure \ref{cosmogic}. We briefly comment on the local density matrix for such an observer in appendix \ref{denmat}.

\subsection{Round sphere example}
\label{sphex}

In this section, we discuss the action \nref{Wliouv} for the concrete case of a sphere, where  $\hat{\gamma}_{b}$ is the metric of a round sphere of radius one. We also choose the usual   bulk de Sitter filling in both bra and ket. Namely, we write the de Sitter metric
\begin{equation}
ds^{2}=-d\tau^{2}+\cosh^{2}\tau\,d\Omega_{d}^{2}=\frac{-d\eta^{2}+(1+\frac{\eta^{2}}{4})^{2}d\Omega_{d}^{2}}{\eta^{2}}
\end{equation}
with $d\Omega_{d}^{2}$ the metric of $S^{d}$, and we introduced the asymptotic conformal time $\eta=-2e^{-\tau}$. We will consider a boundary surface $\Sigma$ which is at a large fixed value of $\tau=\tau_{b}$, or equivalently at small $\eta_{b}$. The no-boundary geometry in the ket consists of the $\tau$ contour going from $\tau=i\pi/2$, where the $S^{d}$ shrinks, to $\tau_{b}$. Different contours parameterizing the same endpoints are just simple contour deformations of the same geometry. The no-boundary geometry in the bra is simply the complex conjugate of the $\tau$ contour with the opposite orientation, so it caps off at $\tau=-i\pi/2$ instead.

We first compute $\mathcal{Q}_{\gamma_{b}}$ by directly solving   \nref{Neq}. Due to the symmetry of the sphere,   $N$  depends only  on $\tau$. One can therefore simply solve \nref{Neq} for the ket and bra solutions $N_{\pm}$, regular at $\tau=\pm i \pi/2$ respectively,  to obtain
\begin{equation}
\label{Nsp+-}
\partial_{\tau}N_{\pm}=-\frac{d}{(\cosh \tau)^{d}}\int_{0}^{\tau} d\tau'\,(\cosh \tau')^{d}\pm \frac{id B_{d}}{(\cosh \tau)^{d}}
\end{equation}
where we split the integral from $\pm i\pi/2$ to $0$, and from $0$ to $\tau$. The first integral is purely imaginary and the second is purely real. Using \nref{Nsp+-} and the definition \nref{norm+-},  after rewriting the imaginary part in terms of $\eta$,  we find 
\begin{equation}
\label{Qsph}
\mathcal{Q}_{S^{d}}=\Gamma(d) ~~~~~\longrightarrow ~~~~~~ \CI_\ell|_{\rm linear} = d Q^2 \zeta_0 ~,~~~~~{\rm with}~~~~\zeta_0 = { 1 \over {\rm Vol}(S^d) } \int d^d x \sqrt{\hat \gamma } \zeta
\end{equation}
where $\zeta_0 $ is the constant mode on the sphere.

We note that the coefficient of the linear term in \nref{Qsph} is related to the normalization of the short distance two-point function \nref{zetatwopt}. This relation follows from conformal covariance and is explained in appendix \ref{SphereCharge}.

We now discuss the operator $\mathcal{P}_{\gamma_{b}}$ for the sphere. For this, we have to discuss the solutions to the wave equation \nref{feq}. After expanding in spherical harmonics of angular momentum $l$, we find 
\begin{equation}
{ d^2 f_{l,\pm} \over d\tau^2} +d \tanh \tau\,{ d f_{l,\pm} \over d \tau } +\frac{l(l+d-1)}{\cosh^{2}\tau}f_{l,\pm}=0
\end{equation}
The solutions with  the proper regularity conditions at $\tau=\pm i\pi/2$ are
\begin{equation}
f_{l,\pm} \propto (\cosh \tau)^{l}{}_{2}F_{1}\bigg(l,l+d,l+\frac{d+1}{2},\frac{1 \pm i \sinh \tau}{2}\bigg)
\end{equation}
where the prefactor of $f_{l,\pm}$ is fixed by requiring $f_{l,\pm}(\tau_{b})=1$. By using the connection formula to expand this function at large values of $\tau$, we obtain the asymptotic expansion
\begin{equation}
f_{l,\pm}=1+ \cdots \mp\frac{iB_{d}}{\Gamma(d)}\frac{\Gamma(l+d)}{\Gamma(l)}((-\eta)^{d}-(-\eta_{b})^{d})+ \cdots
\end{equation}
where the first dotted terms are real, and more leading than $(-\eta_{b})^{d}$. By comparing this expansion to \nref{fdisc}, we can find  the eigenvalue of  $\mathcal{P}_{\gamma_{b}}$ for each  spherical harmonic 
\begin{equation}
\label{Psph}
\mathcal{P}_{S^{d}} \,Y_l = \frac{\Gamma(l+d)}{\Gamma(l)} \,Y_l = l(l+1)\cdots(l+d-1)\, Y_l
\end{equation}
We see that for $l=0$, which corresponds to constant functions, $\mathcal{P}$ is zero as we discuss in section \ref{asygen}. For large $l$, it behaves as $l^{d}$, consistent with the short distance behavior in  \nref{Pflat}. 

In $d=2$, \nref{Psph} is the usual Laplacian $-\nabla^{2}$ on $S^{2}$ and we have  
\begin{equation}
\label{Wliouv2d}
{\cal I}_\ell=\frac{Q^{2}}{4\pi}\int_{S^{2}} d^{2}x\sqrt{\hat{\gamma}_{b}}\bigg[(\hat{\nabla}\zeta)^{2}+2\zeta\bigg]=\frac{Q^{2}}{4\pi}\int_{S^{2}} d^{2}x\sqrt{\hat{\gamma}_{b}}\bigg[(\hat{\nabla}\zeta)^{2}+\hat{R}\zeta\bigg]\qquad (d=2)
\end{equation}
where we used that $\hat{R}=2$ for the round sphere in $d=2$. This is the usual Liouville theory, except for the fact that it is missing the Liouville potential. As a side comment, note that Liouville theory has also appeared in the context of de Sitter gravity in \cite{Collier:2025lux}, which should not be confused with the present discussion. In \cite{Collier:2025lux} a pure gravity theory with no inflaton was considered. By contrast, here we are considering an inflationary theory in three dimensions. 

More generally, the action \nref{Wliouv} on the sphere, using \nref{Qsph} and \nref{Psph}, matches a higher-dimensional generalization of Liouville theory discussed in \cite{Levy:2018bdc, Kislev:2022}, but without the Liouville potential. For even $d$, the action is a local function of $\zeta$, and the operator $\mathcal{P}$ in \nref{Psph} matches the so-called critical GJMS operator \cite{Graham:1992gjms} on the sphere. In that case, \nref{Qsph} also matches the so-called Branson Q-curvature \cite{Branson:1995SharpInequalities}. For more details on that, see section \ref{prevdef}, where we show that the action for $\zeta$ can be expressed in terms of the critical GJMS operator and the Q-curvature for a more general class of geometries. Another comment is that such an action for the random surface using the critical GJMS operator and the Q-curvature was also discussed in \cite{Schiavo:2021unx}, where they propose a generalization of Liouville gravity in higher even $d$. Our setting therefore provides a realization of this proposed random surface theory using the reheating surface in inflation in a particular linear roll limit \nref{LinRol}. 
For the particular case of $d=4$, we have \cite{Levy:2018bdc,Cercle:2019jxx,Schiavo:2021unx}
\be 
{\cal I}_\ell = { Q^2 \over 8 \pi^2 }\int_{S^4} d^4x \sqrt{\hat \gamma } \left[ \zeta (\hat{\nabla}^4 -2 \hat{\nabla}^2 ) \zeta + 12 \zeta \right] ~~~ \qquad (d=4)
\ee 

We also comment that in the relevant case of our universe, with $d=3$, the action can be written as
\begin{equation}
{\cal I}_\ell = {3 Q^2 \over 8 \pi^2 }\int_{S^3} d^3x \sqrt{\hat \gamma } \left[ \zeta (-\hat{\nabla}^{2})\sqrt{1-\hat{\nabla}^{2}}\,\zeta +  4\zeta \right]\qquad ~~~~(d=3)
\end{equation}
which is non-local, as expected from \nref{Pflat}, due to the square root term involving $-\hat{\nabla}^{2}$.

\subsection{Action of conformal isometries}
\label{confiso}

Viewing the theory on $\Sigma_d$ as a theory of gravity on its own, we will need to gauge the coordinate transformations. Some of them act on the tensor part of $\hat \gamma$, which we will fix as part of the integral over conformal classes of metrics. Then there are the reparametrizations that only change the scale factor of the metric. Such reparametrizations can be viewed as part of the scalar mode that we are discussing. These are simply the conformal killing vectors of the sphere and they act as 
\be  \la{SpConf}
\delta \zeta = \xi^i \partial_i \zeta +  { 1 \over d } \hat \nabla \cdot \xi 
\ee 
By the way, this means that $\zeta$ is a logarithmic field of dimension zero, and the normalization of the second term fixes the normalization of $\zeta$ so that $e^{2 \zeta} \hat \gamma_{ij} $ transforms as a metric. 

Under \nref{SpConf}, the action then changes as 
\be 
\delta {\cal I}_\ell \propto \int d^d x\,\sqrt{\hat \gamma}\left[2 \zeta {\cal P}(\xi^i \partial_i \zeta) + \frac{2}{d}\zeta {\cal P}(\hat \nabla \cdot \xi) + 2 {\cal Q}\xi^i \partial_i \zeta + \frac{2}{d}{\cal Q}\hat \nabla \cdot \xi\right]
\ee 
Only the special conformal transformations lead to a non-trivial check. In that case,   $\nabla \cdot \xi$ is an $l=1$ function on the sphere, so that we can drop the last term and argue that 
${\cal P} (\hat \nabla \cdot \xi) = d! \, \hat \nabla \cdot \xi$  using \nref{Psph} in the second term, which then cancels against the third term using \nref{Qsph}. We are left then with the first term. This is the same as the conformal variation of just ${\cal P}$ acting on weight zero fields, say 
\be 
{\cal I } \propto \int d^dx \sqrt{\hat \gamma } f {\cal P}_{\hat \gamma } f ~,~~~~~~\delta f = \xi^i \partial_i f ~,~~~~~~\delta I \propto \int   d^dx  \sqrt{\hat \gamma } f {\cal P } \xi^i \partial_i f 
\ee 
But this vanishes for the following reason. The expression is coordinate covariant and Weyl invariant \nref{QPRel}. This implies that it is invariant under a conformal coordinate transformation. One can also check it directly by using the angular momentum harmonics, that $\xi^i$ sends  the $l$-th harmonic to $l \pm 1$ and that the eigenvalues $p_{l}$ of $\mathcal{P}_{\gamma_{b}}$ satisfy $l\,p_{l+1}-(l+d)p_{l}=0$.  

We can also view these symmetries as arising from bulk isometries. In that context the invariance of the full bulk wavefunctional was discussed in \cite{Chen:2024rpx}, where the same limit \nref{LinRol} was studied.  

We want to emphasize that for a Gaussian action to realize the symmetries \nref{SpConf} on the sphere we {\it need} the linear term proportional to ${\cal Q} $ in the action. In other words, even though we originally derived this term for the Hartle Hawking action, we could have rederived it by saying that we have a Gaussian action that is conformally invariant under \nref{SpConf}. Note that \nref{SpConf} is precisely how the scale factor of the metric transforms under a conformal transformation. 

\section{One-loop corrections and the boundary measure}
\label{loops}

At tree level, the inflaton contribution to the no-boundary wavefunction \nref{Wliouv} gives a quadratic and linear action for the conformal factor of the reheating surface.  There are potential corrections from the one-loop factors $A_+A_-$ in \nref{psibraket}, and we must also account carefully for the measure over boundary conditions when computing the norm or cosmological correlators.  Although one-loop effects are usually subleading to the classical action, in the linear-roll limit the tree-level inflaton action remains of order one as $M_{\rm pl} \to \infty$, and so one-loop corrections can enter at the same order. In even spatial dimensions, where there can be a local Weyl anomaly, we will show that the Weyl dependence of the one-loop factors is precisely canceled by that of the inner-product measure.  Moreover, we will explain how the rolling inflaton modifies the action of the residual conformal transformations on the sphere boundary data, and how this impacts the norm of the no-boundary state.

\subsection{Weyl anomalies and slice independence}

To compute the norm, we integrate $|\Psi[\gamma_b,\phi_b]|^2$ over a gauge-fixed set of boundary configurations with the appropriate inner-product measure. Cosmological correlators are computed from the same integral, with the corresponding observables inserted.  Let $\mathcal{S}$ denote a choice of gauge fixing for spatial diffeomorphisms and the time reparameterization symmetry of \nref{weyl1}, and let $D\mu_{\mathcal{S}}[\gamma_b, \phi_b]$ be the corresponding measure including the Faddeev-Popov determinants.  Schematically, the norm is
\begin{align}
\label{ol_norm}
\mathcal{Z} = \int_{\mathcal{S}} D\mu_{\mathcal{S}}[\gamma_b, \phi_b] \sum_{\text{bra},\,\text{ket}} A_+ A_-\,e^{- \CI_{\rm grav} - \CI_{\phi}}
\end{align}
In the reheating gauge, we use the time-reparameterization freedom to set $\phi_b=\phi_r$, with $\zeta$ then describing the scalar fluctuations of the reheating surface. A transformation \nref{weyl1} changes $\phi_b$ and therefore moves us away from the reheating gauge slice to a different representation of the same physical configuration. The norm must be unchanged if we instead use a gauge slice related by \nref{weyl1}, provided that the wavefunctional and boundary measure are transformed consistently. We will gauge fix on to the reheating surface $\phi_b = \phi_r$ and then integrate only over $\zeta$. In \nref{ol_norm} we need to consider a cutoff for the modes; we take it to be a few times the Hubble scale, as in \cite{Creminelli:2008es}. All shorter distances are integrated out and are expected to cancel out (giving a total factor of 1) in the path integral. Note that this cutoff corresponds to a physical distance measured with respect to the physical metric on the reheating surface $g_{b \, i j} = e^{ 2 \zeta } \hat \gamma_{ b \, ij}/(-\eta_b)^2$.
 
In the linear-roll limit, the nonzero-mode fluctuation problem is independent of the classical inflaton profile. The inflaton potential is linear, so fluctuations about any classical solution are those of a free massless scalar with homogeneous Dirichlet boundary conditions, and the scalar decouples from the gravitational fluctuations as $M_{\rm pl} \to \infty$. Accordingly, the nonzero-mode one-loop factors are independent of the boundary value $\phi_b(x)$; residual gauge factors will be treated separately below.

Na\"{i}vely, invariance under \nref{weyl1} would then imply that the one-loop factors are also independent of the scale factor of the boundary metric. There is, however, a subtlety associated with regularization. We define the cutoff at a fixed physical scale, and different values of $\zeta$ therefore correspond to different coordinate cutoffs. As a result, the regulated one-loop factors can acquire a $\zeta$ dependence. As expected from the underlying invariance under \nref{weyl1}, this dependence is canceled by a corresponding dependence of the boundary integration measure. We verify this cancellation explicitly for the $2+1$-dimensional bulk theory below.

Before we do that, let us remark that in time dependent backgrounds, the one loop factors are expected to have a constrained time dependence. Let us illustrate this for the instructive case of a time dependent harmonic oscillator (which could be a mode of a field in de Sitter)
\begin{equation}
L = \frac{1}{2}\,m(t)\dot{q}^2 - \frac{1}{2}\,k(t) q^2\,.
\end{equation}
We write its wavefunction in Gaussian form,
\begin{equation}
\Psi(q,t) = \alpha(t)\,e^{-\frac{1}{2}C(t)q^2}\,.
\end{equation}
The Schr\"{o}dinger equation gives
\begin{equation}
C = -i m(t)\frac{\dot u}{u}\,,\quad \alpha = \frac{\alpha_0}{\sqrt{u}}\,,
\end{equation}
where $u(t)$ satisfies the equation of motion $\frac{d}{dt}\left(m(t)\dot u\right) + k(t)u = 0$. The Wronskian
\begin{equation}
\mathcal{W} = i m(t)\left(\dot u^{*}u - \dot u u^{*}\right)
\end{equation}
is independent of $t$, and satisfies
\begin{equation}
2\,\text{Re}(C) = \frac{\mathcal{W}}{|u|^2}\,.
\end{equation}
Therefore the norm
\begin{equation}
\int dq\,|\Psi(q,t)|^2 = \frac{|\alpha_0|^2}{|u|}\sqrt{\frac{\pi}{\text{Re}(C)}} = |\alpha_0|^2\sqrt{\frac{2\pi}{\mathcal{W}}}
\end{equation}
is independent of $t$, for a normalizable Gaussian mode on the Lorentzian evolution segment. Thus the time dependence of the Gaussian width is canceled, mode by mode, by that of the prefactor. This gives a direct mode-by-mode check of slice independence for the quadratic fluctuation problem. This discussion is most relevant for the modes shorter than the Hubble scale at the slice. The square of the wavefunction for modes with wavelength larger than Hubble are already time independent. However, this argument shows that if the mode has unit normalization at early times, it should also have a unit normalization at late times. Therefore, for each mode the one loop factor is precisely what is needed so that the integral of $|\Psi|^2$ gives one.

In field theory, the corresponding product over modes requires regularization. A regulator need not preserve the Weyl transformation associated with changing the evaluation surface, so the one-loop factor and the functional measure can separately acquire Weyl dependence even though their product must remain slice independent. We now characterize this dependence under \nref{weyl1}. Define $\mathcal{A}_{d}[\omega;\gamma_b]$ by
\begin{equation}
\label{loopanomaly}
(A_+ A_-)[e^{2\omega}\gamma_b] = e^{\mathcal{A}_{d}[\omega;\gamma_b]} (A_+ A_-)[\gamma_b]\,.
\end{equation}
Notice that in this equation the Weyl factor dependence arises through the dependence of the cutoff procedure on the physical metric. The physical cutoff is defined with the metrics inside the brackets. 

Since the classical weight $\CI$ is invariant under \nref{weyl1}, the product of the one-loop factor and the boundary measure must also be invariant. Consequently, if the one-loop factor transforms as in \nref{loopanomaly}, then, after expressing the two gauge-fixed integrals in the same variables, the boundary measure must transform with the inverse factor. Denoting by $\mathcal{S}_\omega$ the gauge slice obtained from $\mathcal{S}$ by the transformation \nref{weyl1}, we have
\begin{equation}
\label{measureanomaly}
D\mu_{\mathcal S_\omega} = e^{-\mathcal{A}_{d}[\omega;\gamma_b]} D\mu_{\mathcal S}\,.
\end{equation}
We see that the one-loop factor and the boundary measure can each depend on the choice of evaluation surface, while their product does not. Consistency under successive changes of the evaluation surface further requires $\mathcal{A}_{d}$ to obey
\begin{equation}
\label{anomalyWZ}
\mathcal{A}_{d}[\omega_1 + \omega_2; \gamma_b] = \mathcal{A}_{d}[\omega_1; e^{2\omega_2} \gamma_b] + \mathcal{A}_{d}[\omega_2; \gamma_b]\,.
\end{equation}
This is the Wess-Zumino consistency condition for the Weyl anomaly. It should not be confused with the separate redundancy \nref{Gsym} in the decomposition $\gamma_b = e^{2\zeta} \hat{\gamma}_b$, although the logarithmic gravity action obeys an analogous relation under that transformation. We will verify the cancellation explicitly in $2+1$ dimensions below.

The final conclusion of this discussion is that we do not expect any extra $\zeta$ dependence from the combined effects of the measure and the one loop factors. In particular, this also implies that we can change from a physical cutoff defined with respect to the full physical metric to a fiducial cutoff defined with the metric $\hat \gamma_b$ and we will not change the action for $\zeta$.\footnote{This last point is definitely {\it not} the case in usual $d = 2$ Liouville theory \cite{Distler:1988jt, David:1988hj}, because in that case there is no analog of the one loop factors  $A_{+}A_{-}$, only the contribution in the measure.} Note that a particular gauge-fixed presentation may still contain field-dependent factors associated with fixing the residual conformal transformations. We view these factors as part of the gauge-fixed measure, as discussed in section \ref{nbnonzero}. 

\subsection{Explicit check in $2+1$ dimensions}

We now demonstrate the cancellation described above explicitly in $2+1$ dimensions. In two boundary dimensions the local Weyl anomaly is characterized by the central charge. We first use the one-loop determinants on $H^3$, whose conformal boundary is $S^2$, to determine the local Weyl anomaly of the bra-ket factor $A_+ A_-$. Since the anomaly coefficient is local, the result applies to a boundary Riemann surface of arbitrary topology. We then compute the Weyl dependence of the gauge-fixed boundary measure on a general closed Riemann surface, including the contributions from conformal Killing vectors, isometries, and moduli. The one-loop factor and the boundary measure have effective central charges $25$ and $-25$, respectively, so their local Weyl anomalies cancel. For the round sphere, this means that there is no net one-loop shift of the coefficient $Q^2$ in the $d=2$ logarithmic gravity action \nref{Wliouv2d}.  

It is useful to define the Liouville Wess-Zumino functional  
\begin{equation}
S_{\rm L}[\omega;\gamma] = \frac{1}{4\pi}\int d^2 x\,\sqrt{\gamma}\left[(\nabla\omega)^2 + R_{\gamma}\omega\right]\,.
\end{equation}
In two dimensions, we parameterize the Weyl anomaly of a functional $Z[\gamma]$ by an effective central charge $c$, defined through
\begin{equation}
Z[e^{2\omega}\gamma] = e^{\frac{c}{6}S_{\rm L}[\omega;\gamma]} Z[\gamma]\,.
\end{equation}

Let us first consider the one-loop factor. In the linear-roll limit, the bulk fluctuation problem factorizes into the pure-gravity problem and that of a decoupled massless scalar. We only need the Weyl-dependent part of the one-loop determinants that contributes to $A_+ A_-$, rather than their full complex phases. This part can be obtained from the corresponding calculation in $H^3$. The quadratic fluctuation problems on the Hartle-Hawking contour are related by analytic continuation to those of the corresponding $H^3$ problem, with no-boundary regularity continuing to regularity in the interior. For the massless scalar, in particular, the quadratic operators appearing in the Gaussian exponent are related by
\begin{equation}
\mathcal{O}_{{\rm HH},\pm} = \pm i\,\mathcal{O}_{H^3}\,,
\end{equation}
where the two signs correspond to the conjugate ket and bra contours. For a fixed spectral problem and a fixed choice of spectral cut, zeta-function regularization gives
\begin{equation}
\log \det(\pm i\,\mathcal{O}) = \log \det \mathcal{O} \pm \frac{i\pi}{2}\,\zeta_{\mathcal{O}}(0)\,,
\end{equation}
up to the choice of branch. This relation by itself does not establish equality of the full finite-boundary Hartle-Hawking and $H^3$ determinants since the two problems differ globally in their contours and cutoff surfaces. We will use the continuation only to determine the real local Weyl-dependent part of the determinant, with the same renormalization prescription on the two sides. For a conjugate bra-ket pair the associated continuation phases cancel. We use the analogous prescription for the gravitational determinant, including the bulk gravitational ghosts; the resulting real one-loop shift agrees with a direct dS$_3$ analysis~\cite{Cotler:2019nbi}.

We can therefore use the corresponding $H^3$ determinants to determine the Weyl dependence of the one-loop factors. Write the $H^3$ metric as
\begin{equation}
ds_{H^3}^2 = d\rho^2 + \sinh^2\rho\,d\Omega_2^2\,,
\end{equation}
where $\rho$ is the radial coordinate, with the conformal boundary at $\rho \to \infty$. A local Weyl transformation $\gamma \to e^{2\omega}\gamma$ of the boundary metric can be implemented asymptotically by moving a large radial cutoff surface from $\rho = \rho_0$ to $\rho = \rho_0 + \omega(x)$. If $V_{\rm ren}[\gamma]$ denotes the renormalized volume of the corresponding $H^3$ filling, then
\begin{equation}
\label{Vrentransform}
V_{\rm ren}[e^{2\omega}\gamma] - V_{\rm ren}[\gamma] = -\pi S_{\rm L}[\omega;\gamma]\,.
\end{equation}
The $(\nabla\omega)^2$ term here arises from the counterterms needed to renormalize the wiggly cutoff surface. The volume-dependent part of the one-loop determinants on $H^3$ can be understood directly from the heat kernel representation. For a quadratic fluctuation operator $\mathcal{O}$,
\begin{equation}
\log \det \mathcal{O} = -\int_0^\infty \frac{ds}{s}\,\text{Tr}\,e^{-s\mathcal{O}}\,.
\end{equation}
On complete $H^3$, the coincident heat kernel is position independent. Regulating the trace by restricting the volume integral to the region bounded by $\rho = \rho_0 + \omega(x)$ therefore gives
\begin{equation}
\text{Tr}_{\rm reg}\,e^{-s\mathcal{O}} = \int_{H^3_{\rho_0,\omega}} d^3x\,\sqrt{g}\,K_{\mathcal{O}}(s;x,x) = V_{\rm reg}(\rho_0,\omega)\,K_{\mathcal{O}}(s)\,.
\end{equation}
We see that the volume-dependent part of the regulated logarithm of the determinant is proportional to $V_{\rm reg}$. After subtracting the local divergent counterterms, the corresponding contribution to the renormalized determinant is proportional to $V_{\rm ren}$. It is this renormalized finite part that determines the Weyl anomaly below.

For a massless scalar and for gravity, including the bulk gravitational ghosts, the explicit $H^3$ determinant calculations give~\cite{Giombi:2008vd, Cotler:2018zff}
\begin{equation}
\label{Vrencoeff}
\log A_{\phi,+} = \frac{1}{12\pi}\,V_{\rm ren}\,,\qquad
\log A_{{\rm grav},+} = -\frac{13}{6\pi}\,V_{\rm ren}\,,
\end{equation}
up to Weyl-independent terms and phases. We review the derivation of these coefficients in appendix~\ref{app:H3det}. Here $A_{\phi,+}$ and $A_{{\rm grav},+}$ denote the scalar and gravitational one-loop factors for a single ket saddle, so that
\begin{equation}
A_+ = A_{\phi,+} \, A_{{\rm grav},+}\,.
\end{equation}
The corresponding factors $A_{\phi,-}$ and $A_{{\rm grav},-}$ are associated with the bra saddle.

Using the variation of $V_{\rm ren}$ in \nref{Vrentransform} together with \nref{Vrencoeff}, the scalar and gravitational factors for a single ket transform as
\begin{equation}
A_{\phi,+}[e^{2\omega}\gamma] = e^{-\frac{1}{12}S_{\rm L}[\omega;\gamma]} A_{\phi,+}[\gamma]\,,\quad
A_{{\rm grav},+}[e^{2\omega}\gamma] = e^{\frac{13}{6}S_{\rm L}[\omega;\gamma]} A_{{\rm grav},+}[\gamma]\,.
\end{equation}
Thus the scalar and gravitational one-loop factors have effective central charges $-1/2$ and $13$, respectively, and the one-loop prefactor of a single ket has
\begin{equation}
c_{A_+} = -\frac{1}{2} + 13 = \frac{25}{2}\,.
\end{equation}
The bra one-loop prefactor has the same real local Weyl anomaly. Since the local Weyl-dependent contribution is the same on the two sides, we find
\begin{equation}
\label{Atransformation}
(A_+ A_-)[e^{2\omega}\gamma] = e^{\frac{25}{6}S_{\rm L}[\omega;\gamma]}(A_+ A_-)[\gamma]\,,
\end{equation}
and hence
\begin{equation}
c_{A_+ A_-} = 25\,.
\end{equation}
Since the Weyl anomaly is local, the coefficient obtained from the $H^3$ calculation is independent of the topology of the boundary surface. So the same local anomaly governs $A_+ A_-$ on a general Riemann surface, up to Weyl-invariant terms that may depend on the moduli and on the choice of bulk filling.

We now turn to the boundary measure. Its local anomaly can already be anticipated from the usual two-dimensional gravity counting. The scalar boundary measure contributes effective central charge $+1$, and the diffeomorphism ghosts contribute $-26$, giving $c_{\rm measure} = -25$. On a compact surface, one must also carefully treat the finite-dimensional contributions from zero modes, conformal Killing vectors, isometries, and moduli. We will show explicitly that, after accounting for these effects, the measure indeed has $c_{\rm measure} = -25$.

It is useful to consider a constant Weyl transformation, which makes the global scaling and the treatment of the zero modes particularly transparent. Write
\begin{equation}
\gamma_{ab} = L^2 \hat{\gamma}_{ab}\,,
\end{equation}
where $\hat{\gamma}_{ab}$ is a fixed constant-curvature representative and $L$ sets the overall length scale. For $\omega = \log L$, the Gauss-Bonnet theorem gives
\begin{equation}
S_{\rm L}[\log L;\hat{\gamma}] = \chi \log L\,,
\end{equation}
where $\chi = 2 - 2g$ is the Euler characteristic of the boundary surface.  A functional with effective central charge $c$ therefore scales as
\begin{equation}
\label{ZLscaling}
Z[L^2\hat{\gamma}] \sim L^{\frac{c}{6}\,\chi}\,,
\end{equation}
and so the dependence of the gauge-fixed measure on $L$ determines its effective central charge.

At one-loop order around a homogeneous rolling background, we fix the boundary metric to a constant-curvature representative, leaving its moduli unfixed. The scalar is then used to fix the residual combinations of conformal transformations and time reparameterizations that preserve this metric gauge. After appropriate field redefinitions and suppressing factors that are independent of $L$, the gauge-fixed measure takes the schematic form
\begin{equation}
\label{muS}
D\mu_{\mathcal S} \propto D_{\text{moduli},L}g\,D_L'\varphi\,Z_\text{ISO}\,\left(\det{}'\Delta_1\right)^{1/2}\left[\det\left(-\nabla^2|_\text{CKV}\right)\right]^{1/2}\,,
\end{equation}
where $\varphi$ denotes the boundary scalar fluctuation and
\begin{equation}
(\Delta_1)_a{}^b = -\nabla^2\delta_a{}^b - R_a{}^b\,.
\end{equation}
Here $|\text{CKV}|$ denotes the number of conformal Killing vectors which are not isometries, $|\text{ISO}|$ the number of isometries, and $|\text{moduli}|$ the number of real moduli.  Also $-\nabla^2|_{\rm CKV}$ denotes the scalar Laplacian restricted to the scalar functions associated with the non-isometric conformal Killing vectors. The prime on $D_L'\varphi$ indicates that the scalar modes used to fix the residual conformal transformations are omitted, and the prime on the vector determinant indicates that we have omitted its zero modes.

The powers of $L$ in the different factors are as follows. Each real modulus contributes one power of $L$,
\begin{equation}
D_{\text{moduli},L}g \sim L^{|\text{moduli}|}\,.
\end{equation}
The scalar measure has heat kernel coefficient $\chi/6$,\footnote{For a Laplace type operator $\mathcal{O} = -(\nabla^2 + E)$ on a closed two-dimensional manifold, we have $a_2 = (4\pi)^{-1}\int d^2x\,\sqrt{\gamma}\,\text{tr}(E + R/6)$. For a massless scalar, $E = 0$, giving $a_2 = \chi/6$. For the boundary diffeomorphism operator $(\Delta_1)_a{}^b$ we have $E_a{}^b = R_a{}^b$ and so $\text{tr}\,E = R$ and $a_2 = 4\chi/3$.} while omitting the $|\text{CKV}|$ scalar modes used in the gauge fixing contributes an additional factor of $L^{-|\text{CKV}|}$. Thus
\begin{equation}
D_L'\varphi \sim L^{\frac{1}{6} \chi - |\text{CKV}|}\,.
\end{equation}
Dividing by the volume of the isometry group gives
\begin{equation}
Z_\text{ISO} \sim L^{-2|\text{ISO}|}\,.
\end{equation}
Now for the vector determinant, we note that it has heat kernel coefficient $4\chi/3$.  Taking into account the conformal Killing vectors and isometries omitted from the determinant gives
\begin{equation}
\left(\det{}'\Delta_1\right)^{1/2} \sim L^{-\frac{4}{3} \chi + |\text{CKV}| + |\text{ISO}|}\,.
\end{equation}
Lastly, the finite-dimensional Jacobian associated with the conformal Killing vectors contributes
\begin{equation}
\left[\det\!\left(-\nabla^2|_\text{CKV}\right)\right]^{1/2} \sim L^{-|\text{CKV}|}\,.
\end{equation}
Multiplying all of the above factors as per \nref{muS} we land on the overall scaling
\begin{equation}
D\mu_{\mathcal S} \sim L^{-\frac{7}{6}\,\chi + |\text{moduli}| - |\text{CKV}| - |\text{ISO}|}\,.
\end{equation}
But for a closed Riemann surface, we have the relation
\begin{equation}
|\text{CKV}| + |\text{ISO}| - |\text{moduli}| = 3\chi\,,
\end{equation}
and therefore find the nice scaling
\begin{equation}
D\mu_{\mathcal S} \sim L^{-\frac{25}{6} \chi}\,.
\end{equation}
Comparing with the general scaling $L^{\frac{c}{6} \chi}$ in \nref{ZLscaling}, we see that the measure has
\begin{equation}
c_{\rm measure} = -25\,.
\end{equation}

The constant-Weyl scaling determines the coefficient of the local Weyl anomaly of the measure. In two dimensions, using a local covariant regularization, locality and Wess-Zumino consistency then fix the corresponding nontrivial finite Weyl anomaly, up to the Weyl variation of local counterterms. In the renormalization scheme used above, the local anomalous part of the boundary measure therefore transforms as
\begin{equation}
D\mu_{\mathcal S_\omega} = e^{-\frac{25}{6}S_{\rm L}[\omega;\gamma]}D\mu_{\mathcal S}\,.
\end{equation}
We see that this local anomaly is precisely the inverse of the transformation of $A_+ A_-$ found above in \nref{Atransformation}, and so the product entering the norm is Weyl invariant,
\begin{equation}
D\mu_{\mathcal S_\omega}\,(A_+ A_-)[e^{2\omega}\gamma] = D\mu_{\mathcal S}\,(A_+ A_-)[\gamma]\,.
\end{equation}
The cancellation has an important consequence. Taken by itself, the Weyl anomaly of the one-loop prefactor would shift the coefficient $Q^2$ in the $d = 2$ sphere action \nref{Wliouv2d}.  The boundary measure cancels this shift exactly. Thus, there is no net one-loop correction from the local Weyl anomaly, and the coefficient $Q^2$ in the physical probability measure remains precisely its tree-level value.

\subsection{The no-boundary norm with an inflaton is not zero}
\label{nbnonzero}
 
The sphere contribution to the no-boundary norm for gravity coupled to a slowly rolling inflaton differs in an important way from the corresponding result in pure Einstein gravity. Throughout this subsection we work in the linear-roll limit described above, in $d+1$ spacetime dimensions (for $d \geq 2$), so that the reheating surface is an $S^d$. In pure gravity, fixing the boundary metric to be round leaves a residual $SO(d+1,1)$ symmetry. The $SO(d+1)$ rotations form a compact subgroup, while the remaining $d+1$ generators are noncompact. For the round Hartle-Hawking saddle in pure gravity, these transformations leave the saddle fixed. Dividing by the volume of this residual noncompact group then gives a factor
\begin{equation} \la{PureGraFa}
\frac{1}{\text{Vol}(SO(d+1,1))} = 0\,,
\end{equation}
which is the origin of the vanishing sphere contribution to the no-boundary norm discussed in~\cite{Cotler:2025gui, Cotler:2026wlk}.

The presence of the slowly rolling inflaton changes this conclusion. We work directly in the reheating gauge, $\phi_b = \phi_r$, so that the scalar fluctuations of the reheating surface are described by $\zeta$. The relevant residual transformations are the noncompact de Sitter isometries accompanied by a time reparameterization chosen so that $\phi_b$ remains fixed. On the reheating surface, their action is precisely the conformal transformation \nref{SpConf},
\begin{equation}
\delta_{\xi}\zeta = \xi^i\partial_i\zeta + \lambda_{\xi}\,,\quad \lambda_{\xi} = \frac{1}{d}\hat{\nabla}_i\xi^i\,.
\end{equation}
The $SO(d+1)$ rotations have $\lambda_{\xi} = 0$, and the remaining $d+1$ conformal Killing vectors have $\lambda_{\xi}$ in the $l = 1$ scalar sector. Thus, unlike in pure gravity, the noncompact transformations do not leave the homogeneous configuration $\zeta = \zeta_0$ fixed. Instead, they move it along gauge-equivalent configurations whose leading variation is in the $l = 1$ modes of $\zeta$. The stabilizer of the homogeneous gravity-inflaton configuration is therefore only the compact subgroup $SO(d+1)$.

As shown in section~\ref{confiso}, the full linear-roll probability functional is invariant under these transformations. This does not imply that the $l = 1$ harmonics are zero eigenfunctions of the quadratic operator; indeed, \nref{Psph} gives ${\cal P}_{S^d}Y_A = d!\,Y_A$. The symmetry is a symmetry of the full action, including the term linear in $\zeta$, and acts inhomogeneously through \nref{SpConf}. Consequently, invariance of the full action does not require the $l = 1$ harmonics to be zero modes of ${\cal P}$ by themselves.

To make the gauge fixing explicit, let $Y_A$ for $A = 1,...,d+1$ be a real orthonormal basis of $l = 1$ spherical harmonics on $S^d$, satisfying
\begin{equation}
\hat{\nabla}_i\hat{\nabla}_jY_A = -\hat{\gamma}_{ij}Y_A\,,\quad \int_{S^d}d^d x\,\sqrt{\hat{\gamma}}\,Y_A Y_B = \delta_{AB}\,.
\end{equation}
The corresponding noncompact conformal Killing vectors can be chosen as
\begin{equation}
\xi_A^i = \hat{\nabla}^iY_A\,,
\end{equation}
for which $\lambda_{\xi_A} = -Y_A$. Writing $\zeta(x) = \zeta_0 + \sum_{A = 1}^{d+1}\zeta_A Y_A(x) + \cdots$, an infinitesimal transformation generated by $\epsilon_B\xi_B$ gives
\begin{equation}
\delta\zeta_A = -\epsilon_A + \epsilon_B\int_{S^d}d^d x\,\sqrt{\hat{\gamma}}\,Y_A\hat{\nabla}^iY_B\,\partial_i\zeta\,.
\end{equation}
We can therefore fix the noncompact residual transformations along the lines of \cite{Distler:1988jt,Muhlmann:2021clm} by imposing
\begin{equation} \la{GFCon}
\zeta_A = 0\,,\quad A = 1,...,d+1\,,
\end{equation}
and the corresponding Faddeev-Popov matrix is
\begin{equation}
\mathcal{M}_{AB}[\zeta] = \frac{\partial\zeta_A}{\partial\epsilon_B} = -\delta_{AB} + \int_{S^d}d^d x\,\sqrt{\hat{\gamma}}\,Y_A\hat{\nabla}^iY_B\,\partial_i\zeta\,.
\end{equation}
At a homogeneous configuration $\zeta(x) = \zeta_0$ the second term vanishes, and hence
\begin{equation}
\mathcal{M}_{AB} = -\delta_{AB}\,,\qquad |\det\mathcal{M}\,| = 1\,,
\end{equation}
up to the normalization chosen for the conformal Killing generators. The gauge fixing is therefore nondegenerate near homogeneous $\zeta$ configurations. Away from the homogeneous configuration, the Faddeev-Popov determinant is generally field dependent; this dependence is understood to be included in the gauge-fixed measure. For the present purpose, the important point is that the determinant is finite and nonzero near homogeneous configurations.\footnote{A more global gauge choice is $N_A=\int d^dx\,\sqrt{\hat{\gamma}}\,Y_A e^{(d+1)\zeta}=0$, and to leading order around a homogeneous configuration this reduces to \nref{GFCon}. Using the $l=1$ spherical harmonic identities, one finds
$\mathcal{M}_{AB} = \frac{\partial N_A}{\partial\epsilon_B} \propto -\delta_{AB}\int d^dx\,\sqrt{\hat{\gamma}}\,e^{(d+1)\zeta}$ and so $|\det \mathcal{M}\,| > 0$, showing that the Faddeev-Popov matrix is nondegenerate.} The $d+1$ noncompact transformations are removed by the $l = 1$ conditions \nref{GFCon}, and the compact $SO(d+1)$ subgroup remains as a residual gauge group. At the homogeneous round configuration, this subgroup is also the stabilizer of the saddle. Thus, we find that the factor $1/\text{Vol}(SO(d+1,1)) = 0$ in the pure gravity calculation \nref{PureGraFa} is replaced by a finite Faddeev-Popov factor together with $1/\text{Vol}(SO(d+1))$.

After fixing the $l = 1$ gauge directions and integrating the remaining nonzero modes at one loop, the norm contains schematically
\begin{equation}
\frac{1}{\text{Vol}(SO(d+1))}\int d\zeta_0\,Z'_{1\text{-loop}}\,e^{-\mathcal{I}_{\rm cl}(\zeta_0)}\,.
\end{equation}
Here $Z'_{1\text{-loop}}$ denotes the reduced fluctuation factor after gauge fixing the residual boundary conformal transformations. The prime removes the $l=1$ boundary collective coordinates $\zeta_A$.

Using \nref{Qsph}, the classical action for the remaining homogeneous mode is
\begin{equation}\la{ActCons}
\mathcal{I}_{\rm cl}(\zeta_0) = \mathcal{I}_{*} + d Q^2\zeta_0\,,
\end{equation}
where $\mathcal{I}_{*}$ is independent of $\zeta_0$. If this late-time linear-roll expression is extrapolated over the entire zero-mode direction, the $\zeta_0$ integral is formally divergent. This is quite different from the pure Einstein gravity result: there the noncompact residual gauge volume gives a factor $1/\text{Vol}(SO(d+1,1)) = 0$, whereas here those gauge directions have already been fixed and the remaining divergence is an integral over a physical homogeneous mode. The divergence of the $\zeta_0$ integral is therefore not associated with a gauge symmetry. Rather, it appears only when the late-time expression is extrapolated to values of the homogeneous mode for which the reheating surface is no longer parametrically larger than the Hubble scale. In this regime, the asymptotic expression \nref{ActCons} is no longer reliable and must be replaced by the full finite-size no-boundary saddle, which gives a finite action.  This is the familiar phenomenological problem associated with the $\zeta_0$ dependence, namely the preference of the no-boundary state for smaller reheating surfaces. We see no reason to expect the dominant contribution from small reheating surfaces to be precisely canceled by the remaining contributions to the norm.  We leave these issues for future work.

In summary, the rolling inflaton permits a locally nondegenerate fixing of the noncompact residual conformal transformations,
so this residual gauge factor does not by itself force the sphere contribution to vanish. The norm is dominated by the small reheating surface region, which is beyond the regime of validity of our approximations.

\section{Scattering operators and Weyl covariance}
\la{sec:Scat}

In section \ref{nbdinfl}, we defined the response functions $\mathcal{P}_{\gamma_b}$ and $\mathcal{Q}_{\gamma_b}$ directly from problems with boundary conditions imposed on a large but finite surface. Here we introduce closely related scattering problems with boundary conditions imposed at asymptotic infinity. The advantage of this formulation is that the transformation properties of $\mathcal{P}_{\gamma}$ and $\mathcal{Q}_{\gamma}$ under Weyl rescalings of $\gamma$ become particularly transparent. We will first define the asymptotic problem and show that it reproduces the finite-time response functions relevant for the wavefunctional. We will then derive their Weyl transformation properties and relate them, for a special class of geometries, to operators previously studied in the mathematics literature.

\subsection{Complex de Sitter Asymptotics and generalities}
\label{asygen}

In this section, we discuss generalities about no-boundary spacetimes that are solutions to Einstein's equations with a positive cosmological constant. Many of their properties follow from simple analytic continuation of the known properties of solutions with a negative cosmological constant \cite{Skenderis:2002wp}, which we will just state without further elaboration. We discuss here how these properties interplay with imaginary corrections to various Lorentzian quantities. We also define an idealized scattering problem, which we will later, in section \ref{eqwvfct}, relate to the relevant ones discussed in section \ref{nbdinfl}. As in section \ref{gennb}, we consider a no-boundary spacetime whose metric in the late future is of the form
\begin{equation}
\label{asymet}
ds^{2}=\frac{-d\eta^{2}+\gamma_{ij}(\eta)dx^{i}dx^{j}}{(-\eta)^{2}}
\end{equation}
but we also take the spacetime to have an asymptotic spacelike boundary at $\eta=0$. The asymptotic boundary has a fixed renormalized metric $\gamma_{0\,ij}$, defined  by
\begin{equation}
\lim_{\eta \rightarrow0} \eta^{2}g_{\gamma\,ij}(\eta)=\gamma_{ij}(0)=\gamma_{0\,ij}
\end{equation}
with $g_{\gamma\,ij}(\eta)$ the induced metric of the hypersurface at finite $\eta$. 

Einstein's equations with a positive cosmological constant imply an equation of motion for $\gamma_{ij}(\eta)$. As in the AdS problem \cite{deHaro:2000vlm, Skenderis:2002wp}, $\gamma_{ij}(\eta)$ has a power series expansion near $\eta=0$, whose specific form depends on the dimension of spacetime. However, its general form \cite{deHaro:2000vlm} is
\begin{equation}
\label{gammaexp}
\gamma_{ij}(\eta)=\gamma_{0\,ij}+F_{ij}[\eta,\gamma_{0}]+(-\eta)^{d}h_{d\,ij}[\gamma_{0}]+ \cdots 
\end{equation}
where $F_{ij}[\eta,\gamma_{0}]$ is a local function of $\gamma_{0}$ built from a power series in even powers of $\eta$, and possibly terms of order $\eta^{d}\log(-\eta)$, with all terms of lower order in $\eta$ than $\eta^d$. This is a real function in the sense that if $\gamma_{0}$ and $\eta$ are real, so is $F_{ij}[\eta,\gamma_{0}]$. 

The term $h_{d\,ij}$ is the first ``filling dependent'' term in the small $\eta$ expansion. By this we mean that it is not determined locally by the boundary metric $\gamma_0$, but instead depends on how the no-boundary geometry extends into the interior.\footnote{We use the $[\gamma_{0}]$ index in $h_{d\,ij}[\gamma_{0}]$ simply to refer to the boundary condition of the bulk metric.}\footnote{We should emphasize that for even $d$ and some fillings, such as the usual de Sitter filling, the imaginary part of the function $h_{d\,ij}$ is a local function of the metric, but its sign depends on the filling (bra vs ket).} It can therefore have an imaginary part. The first possible imaginary corrections to $\gamma_{ij}(\eta)$ therefore occur at order $\eta^{d}$.
The dots correspond to terms of higher order than $\eta^{d}$, which can generically also be filling dependent. It will be relevant for us later that the Einstein equations constrain the leading filling dependent term to satisfy
\begin{equation}
\label{trhd}
\text{Tr}\,h_{d}[\gamma_{0}]=\gamma_{0}^{ij}h_{d\,ij}[\gamma_{0}]=G[\gamma_{0}]
\end{equation}
where the trace is taken with respect to the zeroth-order metric $\gamma_{0\,ij}$, and $G[\gamma_{0}]$ is a local real function of $\gamma_{0\,ij}$. For even $d$, we have that $G$ is a non-zero function related to the anomaly, and for odd $d$, $G$ is zero. The relevant fact for us is that, for real $\gamma_{0}$ and $\eta$, the imaginary part of $\gamma_{ij}(\eta)$ is of order $\eta^{d}$, but the trace of this imaginary part, with respect to $\gamma_{0\,ij}$, is zero at this order.

In this context, we define two ``scattering problems'', similar to problems previously introduced in the AdS context \cite{ Graham:2003GZ, Fefferman:2002QCurvaturePoincare}. The first is a homogeneous problem for a function $f$ defined with a Dirichlet condition at asymptotic infinity as
\begin{equation}
\label{homprob}
\nabla^{2}f=0~,~~~ f|_{\eta=0}=f_{0} 
\end{equation}
At small $\eta$, $f$ can be approximated by $f_{0}$, and equation \nref{homprob} fixes corrections perturbatively in $\eta$ locally from $f_{0}$ and $\gamma_{0}$. However, coefficients at order $\eta^{d}$ solve the equation of motion in \nref{homprob} at leading order, and are therefore not fixed by the local equation of motion \cite{Graham:2003GZ}. These terms are fixed by the regularity condition of $f$ in the interior, and are therefore filling dependent. There should be no filling dependencies appearing at lower order in $\eta$, since any further filling dependency would come from the metric, whose first filling dependent term is also at order $O(\eta^{d})$.

The expansion of $f$ therefore has the form
\begin{equation}
\label{fasyexp}
f=f_{0}+F_{\text{asy}}[f_{0},\gamma_{0},\eta]+(-\eta)^{d}f_{d}[\gamma_{0}]+ \cdots
\end{equation}
where $F_{\text{asy}}[f_{0},\gamma_{0},\eta]$ is a local function of both $f_{0}$ and $\gamma_{0}$ that is less than $O(1)$ but more leading than $O(\eta^{d})$, and $f_{d}$ is the first filling dependent term. For even $d$, the term $f_{d}$ might be ambiguous, because if $F_{\text{asy}}$ has log terms of the form $\eta^{d}\log  \mu\eta$, then changing $\mu$ would change $f_{d}$ by a real local term \cite{Skenderis:2002wp}. However, these local ambiguities are common to all no-boundary manifolds with a given $\gamma_{0}$. Therefore, the difference between the solutions $f$ to \nref{homprob} for two different no-boundary manifolds with the same $\gamma_{0}$ starts at order $\eta^{d}$, and is free of local ambiguities. Moreover, this difference should be linear in $f_{0}$ since the problem is linear.

Following the definitions in section \ref{nbdinfl}, we can imagine there is a ``ket'' and a ``bra'' manifold, so that for the same boundary conditions
\begin{equation}
\label{Pasydef}
f|_{\text{ket}}-f|_{\text{bra}}=-\frac{2i B_{d}}{\Gamma(d)}(-\eta)^{d}\mathcal{P}_{\gamma_{0}}{f}_{0}+ \cdots 
\end{equation}
where $\mathcal{P}_{\gamma_{0}}$ is a linear operator\footnote{The operator $\mathcal{P}_{\gamma_{0}}$ depends on the choice of manifolds, but we leave the dependence implicit to avoid clutter.}, the prefactor is chosen for later convenience, and the dotted terms are subleading at small $\eta$. Thus the ``scattering problem'' is the following. Given $f_0$ and the bra and ket manifolds, we compute the $(-\eta)^d$ term in \nref{Pasydef}. Note that the definition \nref{Pasydef} is sensible for two generic bra and ket manifolds; we do not assume the bra manifold is the complex conjugate of the ket one, so $\mathcal{P}_{\gamma_{0}}$ can be complex. A relevant feature of the operator $\mathcal{P}_{\gamma_{0}}$ is that for $f_{0}=1$, we have that $f=1$, therefore
\begin{equation}
\label{P1}
\mathcal{P}_{\gamma_{0}}1=0
\end{equation}

Another useful property is that $\mathcal{P}_{\gamma}$ is symmetric under the natural bilinear pairing on the boundary $\Sigma$. One can see this by considering solutions $f$ and $h$ to the problem \nref{homprob} with boundary conditions $f_{0}$ and $h_{0}$. Then, subtracting the identity $\int_{\mathcal{M}}(f\nabla^{2}h-h\nabla^{2}f) =0$ between the ket and bra manifold, integrating by parts, and using \nref{Pasydef}, we find that
\begin{equation}
\int_{\Sigma} d^{d}x\sqrt{\gamma_{0}}f_{0} \mathcal{P}_{\gamma_{0}}h_{0}=\int_{\Sigma}d^{d}x \sqrt{\gamma_{0}}h_{0} \mathcal{P}_{\gamma_{0}}f_{0}
\end{equation}
When the bra and ket manifolds are complex conjugates of each other, the above further implies that $\mathcal{P}_{\gamma_{0}}$ is self-adjoint.

The second problem we consider is a sourced Dirichlet problem for a function $N$ defined as
\begin{equation}
\label{scprob}
\nabla^{2}N=d~,~~~ \lim_{\eta \rightarrow0^{-}}(N-\log(-\eta))=0
\end{equation}

The leading approximation to $N$ at small $\eta$ is $\log(-\eta)$, and equation \nref{scprob} fixes perturbative corrections in $\eta$ that are local in $\gamma_{0}$. However, due to the fact that the left-hand side of \nref{scprob} involves the same Laplacian as in \nref{homprob}, $N$ has filling dependent terms starting at order $\eta^{d}$, for the same reason $f$ did. It is also useful to note that the expansion of $N$ has the form
\begin{equation}
\label{Uasyexp}
N=\log(-\eta)+N_{\text{asy}}[\eta,\gamma_{0}]+(-\eta)^{d}N_{\text{non-loc}}+ \cdots 
\end{equation}
where $N_{\text{asy}}$ is a local function of $\gamma_{0}$ with terms of order at least $\eta^{2}$, but whose all terms are more leading than $\eta^{d}$. The term $N_{\text{non-loc}}$ is the first filling dependency. Similar to the case for $f$, we characterize the difference in filling dependency between the solutions in the bra and ket no-boundary manifolds via
\begin{equation}
\label{Qasydef}
N|_{\text{ket}}-N|_{\text{bra}}=-\frac{2i B_{d}}{\Gamma(d)}(-\eta)^{d}\mathcal{Q}_{\gamma_{0}}+ \cdots 
\end{equation}
where the dotted terms are subleading at small $\eta$. 

\subsection{Relation to problem at finite time}
\label{eqwvfct}

The relevance of the asymptotic problems defined in section \ref{asygen} is that they are simple to work with, but, a priori,  they are not directly relevant to the wavefunctional problem, since the boundary conditions are defined at asymptotic infinity. The relevant problems in the context of the no-boundary wavefunctional instead have a fixed spacelike surface in the far future, at a finite but small $\eta_{b}$. However, we argue that the distinction does not play a role in determining the form of the filling dependencies at leading order.

The basic reason is straightforward. Passing from the asymptotic boundary at $\eta=0$ to a finite boundary at $\eta=\eta_b$ changes the locally determined terms in the small-$\eta$ expansion, but the first filling-dependent response still appears at order $\eta^d$. Imposing the finite-time Dirichlet condition subtracts the value of this response at $\eta=\eta_b$, so that the factor $(-\eta)^d$ in the asymptotic problem is replaced by $(-\eta)^d-(-\eta_b)^d$. The coefficient multiplying it, and hence $\mathcal{P}$ and $\mathcal{Q}$, is unchanged at leading order.

Consider the Dirichlet problem of finding a no-boundary manifold with a metric \nref{asymet}, with $\gamma_{ij}(\eta)$ such that
\be \la{gammafindir}
\gamma_{ij}(\eta_{b})=\gamma_{b\, ij}
\ee
that is, the metric satisfies Dirichlet boundary conditions in Fefferman-Graham gauge at $\eta=\eta_{b}$. This problem can be simplified by noting that, if we know the solution $\gamma_{ij}(\eta)$, we also know an asymptotic $\gamma_{0\,ij}$ obtained by extending it to $\eta=0$. Then, $\gamma_{ij}(\eta)$ can be seen as the expansion \nref{gammaexp} that follows from this value of $\gamma_{0\,ij}$, and we can fix $\gamma_{0\,ij}$ by imposing \nref{gammafindir}. The key then is to note that the filling dependencies in relating $\gamma_{b\,ij}$ to $\gamma_{0\,ij}$ are of order $\eta^{d}$. By inverting \nref{gammaexp} at $\eta=\eta_{b}$, one then sees that $\gamma_{b\,ij}$ is related to $\gamma_{0\,ij}$ as
\begin{equation}
\label{gammarel}
\gamma_{0\,ij}=\gamma_{b\,ij}+F_{\text{inv}\,ij}[\eta_{b},\gamma_{b}]-(-\eta_{b})^{d}h_{d\,ij}(\gamma_{b})+ \cdots 
\end{equation}
with $F_{\text{inv}\,ij}[\eta_{b},\gamma_{b}]$ contain local terms that start at order $O(\eta_{b}^{2})$ but are all more leading than $O(\eta_{b}^{d})$, and $h_{d\,ij}(\gamma_{b})$ is understood to be the coefficient in \nref{gammaexp} of the manifold when we replace $\gamma_{0} \rightarrow \gamma_{b}$. 

We can use this fact to write a perturbative expansion of the metric $\gamma_{ij}(\eta)$ for the Dirichlet problem at small but finite $\eta_{b}$. For this, it is useful to define the notation $O(\eta_{*}^{k})$ in the perturbative expansion, which counts both factors of $\eta$ and $\eta_{b}$ as the same order of perturbation theory. For example, both $\eta_{b}^{3}$ and $\eta_{b}\eta^{2}$ are $O(\eta_{*}^{3})$. Therefore, in the Dirichlet problem at $\eta=\eta_{b}$, the expansion of the metric \nref{gammaexp} has the form
\begin{equation}
\label{gammadirexp}
\gamma_{ij}(\eta)=\gamma_{b\,ij}+F_{\text{fin}\,ij}[\eta,\eta_{b},\gamma_{b}]+((-\eta)^{d}-(-\eta_{b})^{d})h_{d\,ij}(\gamma_{b})+ \cdots 
\end{equation}
where $F_{\text{fin}\,ij}$ is a local function built of $\gamma_{b}$ that vanishes when $\eta=\eta_{b}$, and whose terms are start at order $O(\eta_{*}^{2})$ but are all more leading than $O(\eta_{*}^{d})$. The important feature of \nref{gammadirexp} is that the filling dependent terms are, again, of order $\eta_{*}^{d}$. Note that the term of order $\eta_{*}^{d}$ only changed format due to the fact that the $\gamma_{0\,ij}$ term in \nref{gammaexp} is related to $\gamma_{b\,ij}$ in a filling dependent way at order $\eta_{b}^{d}$ via \nref{gammarel}.

We now discuss the two scattering problems in \nref{homprob} and \nref{scprob}, but for the manifold with a boundary at small but finite $\eta_{b}$. For the source-free problem \nref{homprob}, the generalization is
\begin{equation}
\label{homprobfin}
\nabla^{2}f=0~,~~~ f|_{\eta=\eta_{b}}=f_{b}
\end{equation}
This problem is different from \nref{homprob} because the surface where we fix $f$ is different, and also because the metric we are keeping fixed, $\gamma_{b\,ij}$, is fixed at $\eta=\eta_{b}$ instead of asymptotic infinity. However, using the same trick as before, we can think of extending both the metric and the solution of \nref{homprobfin} to $\eta=0$, in which case the problem reduces to \nref{homprob} for some $\gamma_{0\,ij}$ and $f_{0}$. $\gamma_{0\,ij}$ is fixed by \nref{gammarel}, and $f_{0}$ is fixed from imposing the Dirichlet boundary condition at $\eta=\eta_{b}$. From equations \nref{fasyexp} and \nref{gammarel}, we see that the only filling dependent change of $f$ at order $\eta_{*}^{d}$ is that $f_{0}$ is related to $f_{b}$ via an extra filling dependent term of order $\eta_{b}^{d}$ due to the $f_{d}$ term in \nref{fasyexp}. Therefore, the difference of solutions to \nref{homprobfin} for two no-boundary manifolds with the same $\gamma_{b\,ij}$ and $f_{b}$, as in \nref{Pasydef}, is given by
\begin{equation}
\label{Pfindef}
f|_{\text{ket}}-f|_{\text{bra}}=-\frac{2i B_{d}}{\Gamma(d)}((-\eta)^{d}-(-\eta_{b})^{d})\mathcal{P}_{\gamma_{b}}f_{b}+ \cdots 
\end{equation}
where we replaced $\gamma_{0} \rightarrow \gamma_{b}$ in $\mathcal{P}$ by absorbing the corrections in the subleading terms. Therefore, the operator $\mathcal{P}$ from the asymptotic problem in \nref{Pasydef} is the same as the one for the Dirichlet problem at finite but small $\eta_{b}$. 

To argue the same for $\mathcal{Q}$, consider the sourced problem
\begin{equation}
\label{scprobfin}
\nabla^{2}N=d~~,~~ N|_{\eta=\eta_{b}}=0
\end{equation}
This can be solved by taking $N$ to be given by the solution to \nref{scprob}, with $\gamma_{0\,ij}$ given by \nref{gammarel}, plus an extra term that is annihilated by $\nabla^{2}$. The extra term can be written as $f=-\log(-\eta_{b})+h$, where $h$ satisfies $\nabla^{2}h=0$, since the constant function $\log(-\eta_{b})$ already does. Equation \nref{scprobfin} implies that $h$ satisfies certain Dirichlet boundary conditions at $\eta=\eta_{b}$, where its boundary value is set to be equal to minus the right-hand side of \nref{Uasyexp} with the log term removed. This implies that the boundary value of $h$, $h_{b}$, is of order $\eta_{b}^{2}$, and is a local function of $\gamma_{b}$ up to filling dependencies of order $\eta_{b}^{d}$. The leading filling dependence of $f$ at small $\eta$ is then given precisely by the filling dependence of the boundary value of $h$ at order $\eta_{b}^{d}$. These are by definition the same filling dependencies as those of $N$ in \nref{scprob}, but evaluated at $\eta=\eta_{b}$, and with an overall minus sign. The filling dependence of $N$, at order $\eta_{*}^{d}$, and at finite but small $\eta$, is therefore that coming from $N_{\text{non-loc}}$ in \nref{Uasyexp} minus the same result evaluated at $\eta=\eta_{b}$.

Therefore, the difference of solutions to \nref{scprobfin} for two no-boundary manifolds with the same $\gamma_{b}$ is given by
\begin{equation}
\label{Qfindef}
N|_{\text{ket}}-N|_{\text{bra}}=-\frac{2i B_{d}}{\Gamma(d)}((-\eta)^{d}-(-\eta_{b})^{d})\mathcal{Q}_{\gamma_{b}}+ \cdots
\end{equation}
where we, again, replaced $\gamma_{0}$ by $\gamma_{b}$ at leading order.  We can therefore conclude that, at least as far as the leading filling dependence is concerned, the asymptotic response functions $\mathcal{Q}$ and $\mathcal{P}$ match the finite time ones discussed in section \ref{nbdinfl}. 

\subsection{Change of foliation and Weyl transformations}
\label{folweyl}

We now study how $\mathcal{P}_{\gamma}$ and $\mathcal{Q}_{\gamma}$ vary under Weyl transformations of $\gamma$. This is a variant of an argument in \cite{Fefferman:2002QCurvaturePoincare}, adapted to our context. The simplest way to argue that is to use the asymptotic problem in section \ref{asygen} and consider a different foliation of the manifold. Take a different time coordinate foliation $\tilde{\eta}$ of \nref{asymet}, such that the metric is written as
\begin{equation}
\label{twomet}
ds^{2}=\frac{-d\eta^{2}+\gamma_{ij}(\eta)dx^{i}dx^{j}}{\eta^{2}}=\frac{-d\tilde{\eta}^{2}+\tilde{\gamma}_{ij}(\tilde{\eta})d\tilde{x}^{i}d\tilde{x}^{j}}{\tilde{\eta}^{2}}
\end{equation}

The equivalence of the two metrics implies a differential equation for $\tilde{\eta}$ in terms of $\eta$, as\footnote{Preserving Fefferman–Graham gauge also requires a spatial coordinate change from $x^{i}$ to $\tilde{x}^{i}(x^{j},\eta)$, taken to be the identity at the boundary and suppressed here. However, the filling dependence of $\tilde{x}^{i}(x^{j},\eta)$ enters only at order $\eta^{d+2}$, too late to affect the responses defining $\mathcal P_{\gamma}$ and $\mathcal Q_{\gamma}$ we discuss here.}
\begin{equation}
\label{diffeta}
g^{ab}\nabla_{a}\tilde{\eta}\nabla_{b}\tilde{\eta}=-\tilde{\eta}^{2} \quad \Longrightarrow \quad (\partial_{\eta}\tilde{\eta})^{2}-\gamma^{ij}(\eta)\nabla_{i}\tilde{\eta} \nabla_{j}\tilde{\eta}=\bigg(\frac{\tilde{\eta}}{\eta}\bigg)^{2}
\end{equation}
which should have a unique solution once one fixes the relation between $\tilde{\eta}$ and $\eta$ at small $\eta$. From \nref{diffeta}, at small $\eta$ it must be that $\tilde{\eta}=e^{\omega}\eta + \cdots$ with the dots subleading at small $\eta$, and $\omega$ a function that does not depend on $\eta$. Therefore, the renormalized asymptotic metric as computed from the foliation $\tilde{\eta}$ is
\begin{equation}
\label{weyltrans}
\tilde{\eta}=e^{\omega}\eta+ \cdots \quad \Longrightarrow \quad \tilde{\gamma}_{0\,ij}=\lim_{\tilde{\eta}\rightarrow0}\tilde{\eta}^{2}g_{\tilde{\gamma}\,ij}=e^{2\omega}\lim_{\eta\rightarrow0}\eta^{2}g_{\gamma\,ij}=e^{2\omega}\gamma_{0\,ij}
\end{equation}

Therefore, the foliation $\tilde{\eta}$ describes an asymptotic metric which is in the same conformal class as $\gamma_{0}$, but which is Weyl-rescaled by $e^{2\omega}$, with $\omega$ fixed by the asymptotic relation of $\tilde{\eta}$ and $\eta$. This allows us to discuss $\mathcal{P}_{\gamma}$ and $\mathcal{Q}_{\gamma}$ for Weyl-rescaled versions of $\gamma$. Before we proceed, however, we note that a useful fact of the expansion \nref{diffeta} is that since the filling dependence of $\gamma_{ij}$ is of order $\eta^{d}$, the relation between $\tilde{\eta}$ and $\eta$ is of the form
\begin{equation}
\label{etaexp}
\tilde{\eta}=\eta e^{\omega}(1+H[\gamma_{0},\omega,\eta]+\eta^{d+2}\tilde{H}+ \cdots )
\end{equation}
with $H[\gamma_{0},\omega,\eta]$ a local function of $\gamma_{0}$ and $\omega$ that is less leading than $O(1)$ but more leading than $O(\eta^{d+2})$, and $\tilde{H}$ the first filling dependent term in the expansion. The important point is that the leading filling dependence in relating $\tilde{\eta}$ to $\eta$ is at $O(\eta^{d+2})$. Therefore, the change of foliation does not introduce any filling-dependent contribution at order $\eta^d$, which is precisely the order from which $\mathcal{P}_{\gamma}$ and $\mathcal{Q}_{\gamma}$ are extracted.

Now we discuss the transformation properties of $\mathcal{P}_{\gamma}$. Consider the asymptotic Dirichlet problem defined in \nref{homprob}. The problem is still the same in the two foliations, but $f$ has different expansions due to the change of $\eta$ to $\tilde{\eta}$, and $\gamma_{0}$ to $e^{2\omega}\gamma_{0}$. We can find how $\mathcal{P}_{\gamma_{0}}$ is related to $\mathcal{P}_{e^{2\omega}\gamma_{0}}$ by relating the leading filling dependencies in the two different expansions for $f$. By taking \nref{fasyexp} for the $\tilde{\eta}$ foliation with asymptotic metric $e^{2\omega}\gamma_{0\,ij}$, we see that replacing $\tilde{\eta}$ by $\eta$ via \nref{etaexp} does not give extra filling dependencies at order $\eta^{d}$. However, the term at order $\tilde{\eta}^{d}\approx e^{d\omega}\eta^{d}$ is rescaled when written in terms of $\eta$. This implies that the $\mathcal{P}$ of the two asymptotic problems, defined as in \nref{Pasydef}, are related as
\begin{equation}
\label{Pweyl}
\mathcal{P}_{e^{2\omega}\gamma}=e^{-d\omega}\mathcal{P}_{\gamma}
\end{equation}
as we wanted to show. Moreover, we can see how $\mathcal{Q}_{\gamma}$ transforms by defining a problem as in \nref{scprob} but for $\tilde{\eta}$, via
\begin{equation}
\nabla^{2}\tilde{N}=d~~,~~\lim_{\tilde{\eta} \rightarrow 0}(\tilde{N}-\log (-\tilde{\eta}))=\lim_{\eta \rightarrow 0}(\tilde{N}-\log (-\eta))-\omega=0
\end{equation}
where we used the relation \nref{etaexp} to rewrite the boundary condition for $\tilde{N}$ using $\eta$. We note that $\tilde{N}=N+f$, with $N$ given in \nref{scprob}, plus a field $f$ that satisfies $\nabla^{2}f=0$ and $f|_{\eta=0}=\omega$. This is just a field $f$ defined from the Dirichlet problem in \nref{homprob}, with asymptotic value given by $\omega$. 

One therefore just needs to compare the leading filling dependencies of $\tilde{N}$, expanded as \nref{Uasyexp}, but in terms of the $\tilde{\eta}$ foliation with asymptotic metric $e^{2\omega}\gamma_{0\,ij}$, with those of $N+f$. Replacing $\tilde{\eta}$ by $\eta$ in the expression for $\tilde{N}$ via \nref{etaexp} does not introduce extra filling dependencies, but it rescales the filling dependent term at order $\tilde{\eta}^{d}=e^{d\omega}\eta^{d}$. Using \nref{Qasydef}, we can therefore derive that
\begin{equation}
\label{Qweyl}
\mathcal{Q}_{e^{2\omega}\gamma}=e^{-d\omega}(\mathcal{Q}_{\gamma}+\mathcal{P}_{\gamma}\omega)
\end{equation}
as we wanted to find. 

\subsection{Relation to previous definitions}
\label{prevdef}

Mathematicians have defined objects \cite{Graham:2003GZ, Fefferman:2002QCurvaturePoincare} very similar to the $\mathcal{P}_{\gamma}$ and $\mathcal{Q}_{\gamma}$ that we discussed, in the context of harmonic problems for geometries that are solutions to Einstein's equation with negative cosmological constant. The metric $ds_{a}^{2}$ of these manifolds is of the form
\begin{equation}
\label{adsmet}
ds_{a}^{2}=\frac{dz^{2}+\gamma_{a\,ij}(z)dx^{i}dx^{j}}{z^{2}}
\end{equation}

In that context, one can define two scattering problems \cite{Graham:2003GZ, Fefferman:2002QCurvaturePoincare} for functions $N$ and $f$ via
\begin{equation}
\begin{aligned}
\label{adsprob}
\nabla_{a}^{2}f&=0~~,~~~~~f|_{z=0} =f_{0}\\
\nabla_{a}^{2}N&=-d~~,~~ \lim_{z \rightarrow0} (N-\log z)=0
\end{aligned}
\end{equation}
with $\nabla_{a}^{2}$ defined from the metric \nref{adsmet}. Then, they define scattering responses $\mathsf{P}_{\gamma}$ and $\mathsf{Q}_{\gamma}$ for these problems, in a dimension dependent way. For even $d$, they define
\begin{equation}
\begin{aligned}
\label{evendmath}
f &=f_{0}+ \cdots -(-1)^{\frac{d}{2}}\frac{2 B_{d}}{\pi \Gamma(d)}z^{d}\log z\,\mathsf{P}_{\gamma}f_{0}+ \cdots \\
N &=\log z+ \cdots -(-1)^{\frac{d}{2}}\frac{2 B_{d}}{\pi \Gamma(d)}z^{d}\log z\,\mathsf{Q}_{\gamma}+ \cdots 
\end{aligned}\qquad (d\ \text{even}) .
\end{equation}
where the first dots are local terms polynomial in $z^{2}$ which are less relevant than the leading term, but more relevant than $z^{d} \log z$, and the second dots are subleading filling dependent terms. Note that the definition of the response functions in \nref{evendmath} is quite different in philosophy from ours, since it uses a coefficient of the small $z$ expansion locally determined from the boundary data, such that $\mathsf{P}_{\gamma}$ and $\mathsf{Q}_{\gamma}$ are locally fixed by $\gamma$ and its boundary derivatives. Moreover, it was shown in \cite{Graham:2003GZ} that the operator $\mathsf{P}_{\gamma}$ is the known ``critical GJMS operator'' \cite{Graham:1992gjms}. Moreover, it was shown in  \cite{Fefferman:2002QCurvaturePoincare} that $\mathsf{Q}_{\gamma}$ is also a known local function of $\gamma$, named the Branson Q-curvature \cite{Branson:1985Conformal, Branson:1995SharpInequalities}.

The relation to our cosmological response functions is particularly useful in this even $d$ context. In the AdS problem, the critical GJMS operator and Branson $Q$-curvature appear in the coefficient of a local $z^d\log z$ term. Upon analytic continuation to de Sitter, the logarithm crosses a branch and generates an imaginary contribution at order $(-\eta)^d$. If the bra and ket are complex conjugates and the analytically continued AdS geometry is real, the other filling-dependent terms at this order are real and cancel in the bra-ket difference. The branch contribution isolates precisely the mathematical $\mathsf{P}_{\gamma}$ and $\mathsf{Q}_{\gamma}$.

More specifically, one can consider a ``ket'' de Sitter problem in \nref{asymet}, imagining there is a boundary in the asymptotic future, and at some point in the past we go in a complex direction via $\eta=i z$. In doing so, the metric changes as
\begin{equation}
ds^{2}=\frac{-d\eta^{2}+\gamma_{ij}(\eta)dx^{i}dx^{j}}{(-\eta)^{2}}=-\frac{(dz^{2}+\gamma_{ij}(iz)dx^{i}dx^{j})}{z^{2}}=-ds_{a}^{2}
\end{equation}

Moreover, the problems in \nref{adsprob} become the same\footnote{The problem for $N$ is slightly different since the boundary conditions have an extra factor of $\pm i \pi/2$ depending on how $\eta$ relates to $z$. But this is a constant and does not affect any of our points.} as \nref{homprob} and \nref{scprob} after we realize that since the metrics are flipped, $\nabla^{2}=-\nabla_{a}^{2}$. One can therefore use \nref{evendmath} analytically continued to $z=i(-\eta)$ to recover the solutions we discussed before, as
\begin{equation}
\begin{aligned}
\label{evendmathct}
f|_{\text{ket}} &=f_{0}+ \cdots -\frac{2 B_{d}}{\pi \Gamma(d)}(-\eta)^{d}\log (-\eta)\,\mathsf{P}_{\gamma}f_{0}-\frac{iB_{d}}{ \Gamma(d)}(-\eta)^{d}\,\mathsf{P}_{\gamma}f_{0}+ \cdots \\
N|_{\text{ket}} &=\log z+ \cdots -\frac{2 B_{d}}{\pi \Gamma(d)}(-\eta)^{d}\log (-\eta)\,\mathsf{Q}_{\gamma}-\frac{iB_{d}}{ \Gamma(d)}(-\eta)^{d}\,\mathsf{Q}_{\gamma}+ \cdots 
\end{aligned}\qquad (d\ \text{even}) .
\end{equation}
where the log terms in \nref{evendmathct} led to local responses at order $(-\eta)^{d}$ simply due to the branch we crossed in $\log z$. We would like to use this observation to relate $\mathcal{P}_{\gamma}$ and $\mathcal{Q}_{\gamma}$ we defined in section \ref{asygen} to $\mathsf{P}_{\gamma}$ and $\mathsf{Q}_{\gamma}$. Note, however, that the operators we defined depend on the bra-ket pair under consideration, so one should not expect them to be related in general. Indeed, in the dots in \nref{evendmathct} there are extra filling dependent terms at order $\eta^{d}$ that we have not discussed explicitly. However, if the AdS manifold \nref{adsmet} is real for real $z$, then these filling-dependent terms are also real.

Therefore, we can consider the situation where the bra manifold is the complex conjugate of the ket one. In this case, the functions $f$ in the ket and bra are complex conjugates, and when taking the difference, all the real terms in \nref{evendmathct} cancel. Therefore, in the case that the bra and ket are complex conjugate manifolds that admit a real AdS section \ref{adsmet}, using \nref{Pasydef} and \nref{Qasydef}, we discover that
\begin{equation}
\label{PQmatheq}
\mathcal{P}_{\gamma}=\mathsf{P}_{\gamma}~,~~~ \mathcal{Q}_{\gamma}=\mathsf{Q}_{\gamma}\qquad (\text{if $d$ is even, bra = ket}^{*},\text{real AdS section})
\end{equation}

In this case, they are given by explicit expressions\cite{Graham:2003GZ}   
\begin{equation}
\begin{gathered}
\mathsf{P}_{\gamma}f=-\nabla^{2}f~,~~~~~~~~\mathsf{Q}_{\gamma}=\frac{R}{2}\qquad(d=2)\\
\mathsf{P}_{\gamma}f=(-\nabla^{2})^{2}f+\nabla_{i}\bigg[\bigg(2R^{ij}-\frac{2}{3}R\gamma^{ij}\bigg)\nabla_{j}f\bigg]~,~ \mathsf{Q}_{\gamma}=-\frac{1}{6}\nabla^{2}R-\frac{1}{2}R_{ij}R^{ij}+\frac{1}{6}R^{2}\qquad (d=4)
\end{gathered}
\end{equation}
At $d=2$, the operator $\mathsf{P}_{\gamma}$ is simply minus the Laplacian, and $\mathsf{Q}_{\gamma}$ is a multiple of the Ricci scalar. In $d=4$, the operator $\mathsf{P}_{\gamma}$ is   known as the ``Paneitz operator'' \cite{Paneitz:2008afy}. Another interesting point is that $\mathsf{Q}_{\gamma}$ is proportional to the holographic Weyl anomaly $\langle T_{i}^{i}\rangle$ of Einstein gravity in AdS \cite{Henningson:1998gx, Skenderis:2002wp}, up to total derivative terms \cite{Graham:2003GZ, Graham:2007Qcurvature}.

However, a warning is that there are no other known examples of no-boundary geometries with a real AdS section that respect the KSW criteria \cite{Kontsevich:2021dmb, Witten:2021nzp}, other than the (possibly deformed) sphere boundary with a (possibly deformed) de Sitter filling discussed in section \ref{sphex}. The de Sitter example is, of course, an important one, since it is believed to be the leading contribution to the wavefunctional\footnote{The rationale is that Euclidean de Sitter (i.e.~$S^{d+1}$) is the connected compact saddle of pure gravity with a positive cosmological constant that has the most negative Euclidean action. This follows from Bishop's theorem \cite{Bishop:1963volume}. However, Bishop's theorem does not apply to the complex saddles relevant to the no-boundary state, and the dominance of the sphere saddle in our context is not rigorously proven.}. But the general applicability of \nref{PQmatheq} seems unclear if one believes the KSW criteria.

We should also comment that \cite{Fefferman:2002QCurvaturePoincare} also defines $\mathsf{P}_{\gamma}$ and $\mathsf{Q}_{\gamma}$ at odd $d$. They are defined as the coefficients in
\begin{equation}
\begin{aligned}
\label{odddmath}
f &=f_{0}+ \cdots +(-1)^{\frac{d+1}{2}}\frac{B_{d}}{\Gamma(d)}z^{d}\mathsf{P}_{\gamma}f_{0}+ \cdots \\
N &=\log z+ \cdots +(-1)^{\frac{d+1}{2}}\frac{B_{d}}{\Gamma(d)}z^{d}\mathsf{Q}_{\gamma}+ \cdots 
\end{aligned}\qquad (d\ \text{odd}) .
\end{equation}
which are generally filling dependent, so relating them to our scattering problem does not necessarily give us more computational power. However, for completeness, we can relate the $\mathcal{P}_{\gamma}$ and $\mathcal{Q}_{\gamma}$ we constructed to these definitions as well. Since there are no further filling dependencies at order $\eta^{d}$ in \nref{odddmath}, we can make the relation precise even when the bra-ket pair used to define our operators is not a complex conjugate pair. Indeed, by repeating our argument from before, we learn that $\mathcal{P}_{\gamma}$ is the average of $\mathsf{P}_{\gamma}$ in \nref{odddmath} between the bra and the ket, and the analog holds between $\mathcal{Q}_{\gamma}$ and $\mathsf{Q}_{\gamma}$.

We conclude that the definition we used in the cosmological case is quite similar to the $d$ odd case in AdS \nref{odddmath}, in the sense that we use the $(-\eta)^d$ terms of the expansion.  In our case, we do not need to distinguish between $d$ even and $d$ odd; it turns out that the logarithmic terms that appear in the $d$ even case cancel out when we consider the difference between the bra and ket manifold solutions.

To summarize, the response functions entering the logarithmic gravity action agree between the finite-time wavefunctional problem and the asymptotic scattering problem at the leading filling-dependent order relevant for the late-time probability measure. The asymptotic formulation makes their Weyl covariance manifest, giving \nref{Pweyl} and \nref{Qweyl}. For complex-conjugate bra-ket fillings with a real AdS section, these response functions furthermore reduce in even dimensions to the critical GJMS operator and Branson $Q$-curvature.
 
\section{Gravitational action and zero mode}
\la{sec:GA}

In this section, we point out two consequences of the response function $\mathcal{Q}_{\gamma_b}$. The classical pure-gravity bra-ket weight can itself be written in terms of $\mathcal{Q}_{\gamma_b}$. This relates the gravitational action directly to the zero-mode action of the inflaton. We then use this relation to discuss the preference of the no-boundary state for small reheating surfaces beyond the round sphere saddle, and to connect this behavior with the KSW criterion.

\subsection{No-boundary wavefunctional for pure gravity}
\la{PureGrav}

Similar to the wavefunctional for the inflaton discussed in section \ref{nbdinfl}, we can compute $\CI_{\text{grav}}$ from \nref{GravAct} and \nref{Wdef} in terms of $\mathcal{Q}_{\gamma_{b}}$, in the context where there is a large surface in the future as described in section \ref{eqwvfct}. This is interesting because it relates the action for the scalar and gravitational degrees of freedom. 

The first step is to calculate the volume of the bra-ket spacetime. One can write it in terms of a boundary derivative of the function $N$ from the Dirichlet problem in \nref{scprobfin}, by using the following identity
\begin{equation}
\begin{aligned}
V_{L}=V_{\text{ket}}-V_{\text{bra}} &=\frac{1}{d}\bigg[\bigg(\int_{\mathcal{M}} d^{d+1}x\,\sqrt{-g}\nabla^{2}N\bigg)_{\text{ket}}-\bigg(\int_{\mathcal{M}} d^{d+1}x\,\sqrt{-g}\nabla^{2}N\bigg)_{\text{bra}}\bigg]\\
&=-\frac{1}{d}\int_{\Sigma} d^{d}x\,(-\eta_{b})^{-d+
1}\sqrt{\gamma_{b}}(\partial_{\eta}N|_{\text{ket}}-\partial_{\eta}N|_{\text{bra}})
\end{aligned}
\end{equation}
where we used the equation of motion for $N$ and integrated by parts. Note that at the level of deriving this formula, this would be true for any function $N$ satisfying $\nabla^{2}N=d$ and regularity conditions; we just picked the problem \nref{scprobfin} for definiteness. Therefore, using \nref{Qfindef}, we derive that, up to subleading terms at small $\eta_{b}$
\begin{equation}
\label{VlQ}
V_{L}=-\frac{2iB_{d}}{\Gamma(d)}\int_{\Sigma}d^{d}x\,\sqrt{\gamma_{b}}\mathcal{Q}_{\gamma_{b}}+ \cdots 
\end{equation}
We see that the leading filling-dependent part of the bra-ket volume is completely encoded in $\mathcal{Q}_{\gamma_b}$. This is the same response function that controlled the term linear in $\zeta$ in the inflaton action \nref{WActL}.

Moreover, \nref{VlQ} is invariant under Weyl transformations of $\gamma_{b}$. This is because under Weyl, $\mathcal{Q}_{\gamma}$ changes as \nref{Qweyl}, so
\begin{equation}
V_{L}|_{e^{2\omega}\gamma_{b}}=-\frac{2iB_{d}}{\Gamma(d)}\int_{\Sigma}d^{d}x\,e^{d\omega}\sqrt{\gamma_{b}}\mathcal{Q}_{e^{2\omega}\gamma_{b}}=-\frac{2iB_{d}}{\Gamma(d)}\int_{\Sigma}d^{d}x\,\sqrt{\gamma_{b}}(\mathcal{Q}_{\gamma_{b}}+\mathcal{P}_{\gamma_{b}}\omega)=V_{L}|_{\gamma_{b}}
\end{equation}
where we omitted subleading terms and used the fact that, after integrating by parts, the $\mathcal{P}_{\gamma_{b}}$ term vanishes due to \nref{P1}. As a side comment, note that one can see \nref{VlQ} as more naturally giving the amount of ``Euclidean volume'' in the no-boundary bra-ket manifold, given by
\begin{equation}
\label{VeQ}
V_{E}=iV_{L}=\frac{2B_{d}}{\Gamma(d)}\int_{\Sigma}d^{d}x\,\sqrt{\gamma_{b}}\mathcal{Q}_{\gamma_{b}}
\end{equation}

For general bra-ket pairs \nref{VeQ} is complex, so with Euclidean volume we just mean we are integrating $\sqrt{g}$ over the manifold instead of $\sqrt{-g}$. However, if the bra and ket are complex conjugates, then the imaginary ``Lorentzian'' evolution in the bra and ket cancel, in which case the net volume is indeed completely ``Euclidean'' and \nref{VeQ} is real. This is indeed what follows from \nref{VeQ}, since in this context $\mathcal{Q}_{\gamma_{b}}$ is real. 

Let us now consider the extrinsic curvature term in \nref{GravAct}. A finite bra-ket difference from this term could arise only from the filling-dependent part of the induced volume element at order $\eta_*^d$. Using \nref{gammadirexp}, we have
\begin{equation}
\sqrt{\gamma(\eta)}=\sqrt{\gamma_{b}}\bigg(1+ \cdots +\frac{1}{2}\big((-\eta)^{d}-(-\eta_{b})^{d}\big)\text{Tr}\, h_{d}(\gamma_{b})+ \cdots \bigg) \,.
\end{equation}
The first dots denote locally determined terms that are more leading than $\eta^d$, while the second dots are subleading to order $\eta_*^d$. The potentially finite contribution is therefore controlled by $\text{Tr}\,h_d$. However, the Einstein constraint \nref{trhd} fixes this trace to be a local function of the boundary metric $\gamma_b$. It is consequently the same for the bra and ket fillings and cancels in their difference. Thus the extrinsic curvature term gives no finite filling-dependent contribution at this order, and only the volume term contributes to the leading bra-ket gravitational weight.\footnote{A similar point was discussed in \cite{Hertog:2024nbh} using \cite{deHaro:2000vlm}. This same argument shows that the AdS renormalized action is given by the renormalized volume, with no finite contribution from the extrinsic curvature.} This also explains why the same leading result is obtained whether the boundary conditions are imposed at asymptotic infinity or at a finite but late surface $\eta_b$.

The classical  gravitational weight  $e^{-\CI_{\text{grav}}}$ can therefore be succinctly written in terms of $\mathcal{Q}$  
\begin{equation}
\label{gravwg}
\CI_{\text{grav}}=-\frac{2dB_{d}}{\Gamma(d)}\bigg(\frac{M_{\rm pl}}{H}\bigg)^{d-1}\int_{\Sigma}d^{d}x\sqrt{\gamma_{b}} \mathcal{Q}_{\gamma_{b}}~,
\end{equation}
where we restored the Hubble constant $H$ for completeness. 

 We also note that for $d=2$, and for complex conjugate bra-ket fillings with a real AdS section as discussed in section \ref{prevdef}, the action \nref{gravwg} becomes proportional to the Euler characteristic $\chi(\Sigma)$ of $\Sigma$ since $\mathcal{Q}_{\gamma_{b}}$ for these fillings is proportional to the Ricci scalar of $\Sigma$. This is interesting because it means that for these saddles the gravitational action acts as a topological suppression in the sum over geometries of the universe at reheating, in the same form that one would obtain in string theory. More precisely,
\begin{equation}
\label{2dliouvgrav}
\CI_{\text{grav}}=-\frac{1}{2}\chi(\Sigma)S_{\text{dS}}~~,~~S_{\rm dS}=\frac{4\pi^{2} M_{\rm pl}}{H}~~(d=2, \text{bra = ket}^{*}, \text{real AdS section})
\end{equation}
which corresponds to $g_{s} \sim e^{-\frac{1}{2}S_{\text{dS}}}$.\footnote{The same Euler-characteristic dependence appears in the inflaton zero mode. In $d=2$, the term linear in a constant $\zeta_0$ is proportional to $\chi(\Sigma)\zeta_0$. Thus it reverses sign for $g>1$, so these higher-genus saddles do not have the sphere's preference for small reheating surfaces. However, the real-AdS-section saddles with $g>0$ considered here are both very suppressed gravitationally and do not satisfy the KSW criteria.} We should note that in \cite{Collier:2025lux} the authors discussed a different wavefunctional of the universe for pure gravity in $d=2$, which also had a similar topological expansion structure to \nref{2dliouvgrav}, and was related to a dual matrix integral. One might therefore wonder if the no-boundary state of pure gravity in $d=2$, with a sum over bulk saddles restricted to those with a real AdS section, also has such an interpretation. We should comment, however, that for contributions where the bra and ket are not complex conjugates, we expect extra phase terms in \nref{2dliouvgrav}. One should also worry that the special bulk saddles we are discussing, for boundaries with $g>0$, do {\it not} respect the KSW criteria \cite{Kontsevich:2021dmb, Witten:2021nzp}. 

\subsection{KSW and the $l=0$ problem beyond the sphere}

We now use \nref{gravwg} to extend the familiar sphere zero-mode problem to more general no-boundary fillings. The main observation is that both the pure gravity action and the term linear in the homogeneous mode $\zeta_0$ are proportional to the same quantity, the bra-ket Euclidean volume \nref{VeQ}. The KSW criterion then constrains the sign of this common factor.

Note that the gravitational action \nref{gravwg} is related to the zero mode action of the inflaton we derived in section \ref{nbdinfl}. To see that, we set $\zeta$ there to a constant $\zeta_{0}$, and note that
\begin{equation}
\label{Wksw}
\CI=\CI_{\text{grav}}+\CI_{\phi}=-\frac{2dB_{d}}{\Gamma(d)}\bigg(\frac{M_{\rm pl}}{H}\bigg)^{d-1}\bigg[1-(d-1)\epsilon \,\zeta_{0}\bigg]\int_{\Sigma} d^{d}x\,\sqrt{\gamma_{b}}\mathcal{Q}_{\gamma_{b}}
\end{equation}
with $\epsilon$ the usual slow-roll parameter, which we defined in equation \nref{epsdef}. Moreover, due to \nref{VeQ}, the two contributions, and thus the whole expression, are proportional to the overall Euclidean volume \nref{VeQ} of the bra-ket no-boundary manifold. As noted in \cite{Hertog:2024nbh}, the KSW criteria \cite{Kontsevich:2021dmb, Witten:2021nzp} imply that the real part of the Euclidean volume of allowable no-boundary manifolds is positive. From \nref{Wksw}, the KSW criteria thus imply, in particular, that in the linear roll limit of inflation the reheating surface wants to be as small as possible, which is the usual problematic prediction of the no-boundary state \cite{Maldacena:2024uhs}. This problem is generally discussed in the context of the sphere saddle, but as we just argued, it is a more general consequence of KSW. Therefore, even if for some reason we were supposed to discard the sphere saddle, we would still have to deal with the issue in other, previously subleading, saddles. 

We also note that there is a simple explanation for the relation between the action for the zero mode of $\zeta$ and the gravitational action, which implies a relation like \nref{Wksw} for more general slow-roll inflation. To argue for that, we now generalize an argument in section 4 of \cite{Maldacena:2024uhs}. Namely, in the context of slow-roll inflation, one can imagine that we still consider a no-boundary geometry with a large universe at reheating. The geometry is quasi-de Sitter, but with a local Hubble $H(\phi)$ fixed by the local value of the scalar. The geometry at late times is mostly Lorentzian, so the late-time region contributes to the action with an overall phase to $\Psi$. However, at early times the geometry becomes considerably complex, and then caps off relatively quickly in a region that is Euclidean in a relevant way (see Figure \nref{hhfig}). We denote by $\phi_{*}$ the value of the inflaton in this early cap region, and approximate $\phi \approx \phi_{*}$ there as constant. We also approximate $\CI$ in \nref{psibraket} by the gravity action in this region. In that case, we can approximate $\CI$ using the gravity contribution in \nref{gravwg}, but with the Hubble constant $H$ evaluated at $\phi = \phi_{*}$,
\begin{equation}
\label{Wwkb}
\CI\approx -\frac{2dB_{d}}{\Gamma(d)}\bigg(\frac{M_{\rm pl}}{H(\phi_{*})}\bigg)^{d-1}\int_{\Sigma}d^{d}x\sqrt{\gamma_{b}} \mathcal{Q}_{\gamma_{b}} ~.
\end{equation}

We can now ask how $\CI$ changes when the reheating surface is made larger by an overall factor $e^{\Delta\zeta}$. This corresponds to moving the early cap approximately $\Delta\zeta$ e-folds backward along the slow-roll trajectory. Using the slow-roll relation $d\phi/dN_{e} \simeq -\partial_{\phi}V/(dH^{2})$, with $N_{e}$ the number of e-folds, moving backward along the trajectory gives
\begin{equation}
\Delta \phi_{*}=\frac{\partial_{\phi}V_{*}}{d H_{*}^{2}}\Delta \zeta
\end{equation}
with $*$ standing for the fact that we are evaluating these quantities at $\phi=\phi_{*}$. Using the definition of the slow-roll parameter for a more general potential, we can thus derive that
\begin{equation}
\Delta \CI\approx(d-1)\epsilon_{*}\Delta \zeta \times \frac{2dB_{d}}{\Gamma(d)}\bigg(\frac{M_{\rm pl}}{H(\phi_{*})}\bigg)^{d-1}\int_{\Sigma}d^{d}x\sqrt{\gamma_{b}} \mathcal{Q}_{\gamma_{b}}
\end{equation}
which has the same dependence on $\zeta$ as \nref{Wksw}, but with $\epsilon$ and all parameters evaluated when the geometry becomes complex in the past. The apparent zero of the linearized expression \nref{Wksw} occurs only when $\zeta_0 \sim 1/\epsilon$, precisely where that approximation breaks down. In this regime one should instead use the finite expression \nref{Wwkb}; the sign of the classical action cannot be reversed by extrapolating \nref{Wksw} beyond its regime of validity.
Of course, in the limit \nref{LinRol}, we have $\epsilon \to 0$. 
 
\subsection{Adding the Liouville potential}

\label{LiouvillePotential}

In this paper, we have found that the probabilities for the geometry of the reheating surface in linear roll inflation are given by a logarithmic gravity action. Out of curiosity,  we would like to ask whether we could turn that into the more conventional Liouville gravity action, which also contains a ``potential'' term of the form $\mu \int \sqrt{\hat \gamma}   e^{ d \zeta } $.

\begin{figure}[t!]
    \centering
    \includegraphics[width=0.7\linewidth]{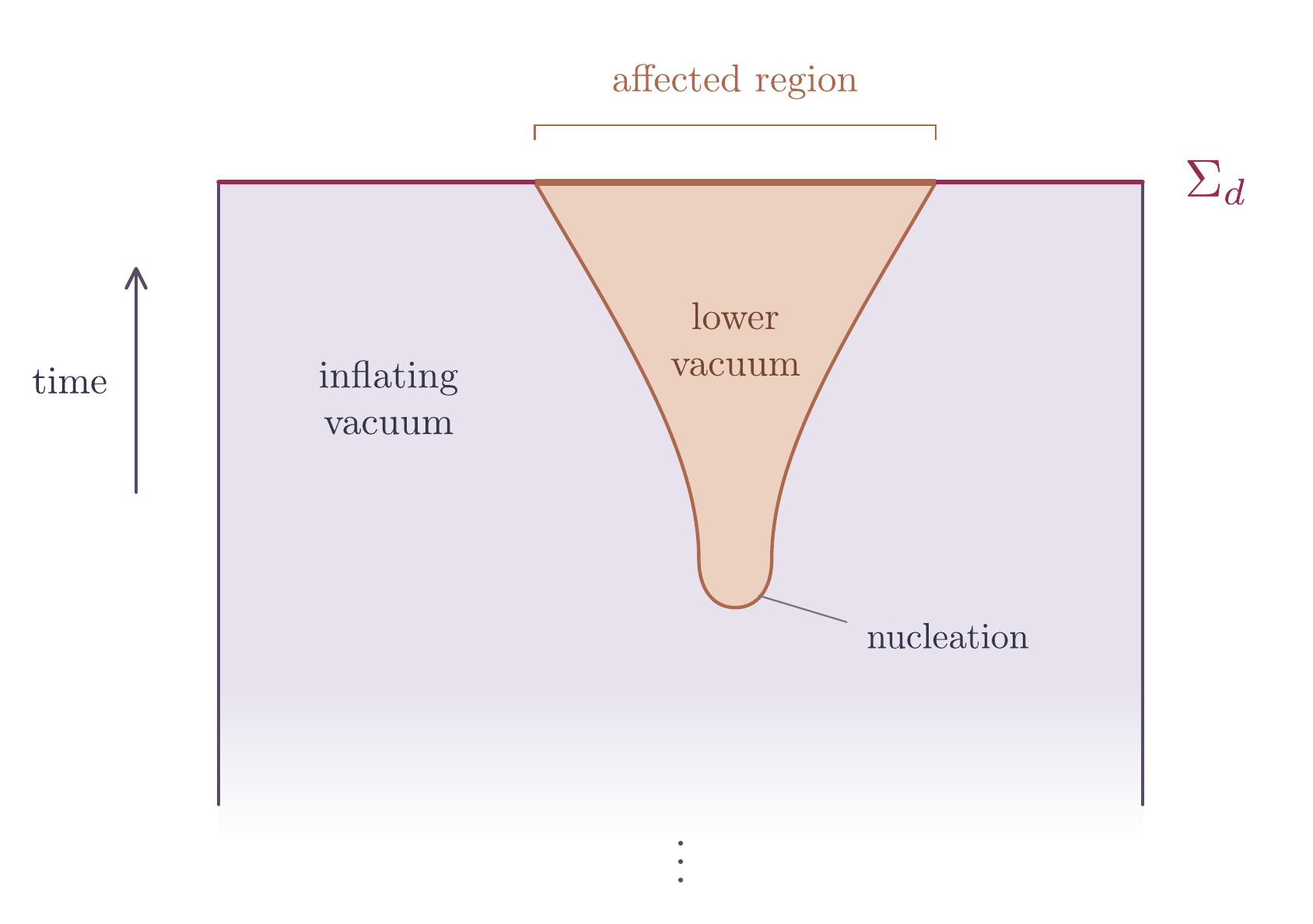}
    \caption{Spacetime where a bubble of a lower vacuum was nucleated and disturbed the geometry of the reheating surface in the future.}
    \label{vacdec}
\end{figure}

This potential term does not arise in the physical situation we have been considering so far. However, we can modify the problem so that this term arises. For that purpose, we can consider a situation where the vacuum of the theory during inflation can locally decay via a bubble nucleation process \cite{Coleman:1980aw}. In other words, we imagine that the almost de Sitter vacuum can decay into a vacuum with a lower cosmological constant. Let us denote by $\Gamma $ the decay rate per unit volume. 
When the vacuum decays, it produces a bubble of the new vacuum. This bubble expands and produces a region inside the reheating surface which is substantially different, see Figure \nref{vacdec}. 

We will now consider the probability that the vacuum has \textit{not} decayed anywhere on the reheating surface, so that no bubbles were nucleated (even though they {\it could} be nucleated).  This probability will be suppressed by a term proportional to the spacetime volume of the manifold\footnote{Here, by spacetime volume we mean the large real part of the Lorentzian volume of the ket manifold. This is different from the bra-ket volume discussed in section \ref{PureGrav}, which was a finite difference of volumes in the bra and ket due to their opposite orientation.} 
\be \la{ActLiou}
 \exp( - \Gamma V_L )  ~,~~~~V_L = { 1 \over H^{d+1} } \int   {d \eta \over (-\eta)^{d+1 } } \,d^d x\,\sqrt{\gamma}   \propto \int d^d x \sqrt{\hat \gamma_b} \,e^{ d \zeta } 
 \ee 
 where the Lorentzian volume is dominated by contributions near the boundary.  For that reason, it is proportional to the physical volume of the boundary surface, up to factors of order $H$.  Here we are imagining that $\Gamma$ stays constant in the $M_{\rm pl} \to \infty $ limit \nref{LinRol}. This could happen if there is an additional field theory with a metastable vacuum which can decay to a more stable one. This could be a field theory in addition to the one describing the inflaton field $\phi$. It could be a field theory which is gapped around the original metastable vacuum. In addition, we could assume that inflation lasts for a long time, in that case we want $\Gamma e^{ d N_{e} }$ to be finite where $N_{e}$ is the number of e-folds, which in the above form of the action we assumed to be included in $\zeta$. Alternatively, we could define the metric in terms of $e^{ 2 \zeta}/\eta_b^2$ and say that $\Gamma/(-\eta_b)^d $ remains finite. 

The volume term in \nref{ActLiou} was written with a physical short distance cutoff. As explained in \cite{Cotler:2026lna}, for $Q>2$  the exponent in \nref{ActLiou} should be changed as 
\be \la{LiouvRen}
 \exp\!\left(  d \zeta \right) ~~~\longrightarrow ~~~  \exp\!\left(  db Q  \zeta \right) ~,~~~~~~~ {\rm with}~~~~~~b= { Q \over 2 } - \sqrt{{  Q^2 \over 4 } -1}
\ee 
when we renormalized the interaction using a fiducial or comoving cutoff.  For $Q<2$ this renormalization breaks down, as one can see from \nref{LiouvRen} since $b$ becomes complex. This corresponds to the regime of eternal inflation \cite{Cotler:2026lna}. 

After these steps, we can say that the probability measure for the shape of a universe that has not yet decayed by the time it gets to the reheating surface is given by the Liouville functional 
\be 
 \exp\left[   -   C_d Q^2 \int \sqrt{\hat \gamma_b} ( \zeta {\cal P }_{\hat \gamma_b} \zeta + 2 \zeta {\cal Q }_{\hat \gamma_b}   + \mu e^{ d Q b \zeta }  ) \right] ~,~~~~~~~~\mu \propto { \Gamma \over H^{d+1} }{ 1 \over (-\eta_b)^{d}}
 \ee 
 with $C_d $ the numerical constant given in \nref{Wliouv}.  

 The Liouville potential suppresses large volumes. This is for a simple reason: the larger the volume, the smaller the probability of {\it not} creating any bubble. 

One amusing observation is the following. The Liouville potential is like a cosmological term in the Liouville gravity theory. From the Euclidean gravity point of view, it is a term that can be generically present, and is expected to be, since it is generated under renormalization. However, in our non-decaying linear roll setup, we have seen that it was not generated by one loop effects.
 The physical reason is that, unlike in usual field theory, integrating out short-distance modes in our context simply produces factors of $1$, since the integrated norm is computing probabilities in a unitary theory, as discussed in section \ref{loops}. The usual renormalization mechanism that produces a cosmological constant is therefore trivial here. So in some sense, the cosmological term appears to be ``forbidden'' by bulk unitarity. This seems to be interesting both as a physical situation where a cosmological constant term is not generated, and also as an example of a term that is absent in the boundary CFT by bulk unitarity.  In the present discussion, we emphasize that the boundary cosmological constant is only allowed because we effectively violate unitarity by post-selecting on configurations where the vacuum has not decayed.

We should point out that the cosmological constant term does not alleviate the familiar Hartle-Hawking preference for small reheating surfaces reviewed in \cite{Maldacena:2024uhs}. The cosmological constant term suppresses large reheating volumes, and therefore reinforces rather than compensates this preference.  

It would be interesting to know whether a $d$-dimensional version of the DOZZ formula \cite{Dorn:1994xn, Zamolodchikov:1995aa, Teschner:1995yf} exists.\footnote{This question was investigated in \cite{Furlan:2018jlv}, and built upon by \cite{Kislev:2022}. Unfortunately, \cite{Furlan:2018jlv} was withdrawn.}

\subsubsection{Attempt to balance the volume operator against the sphere pressure}

As a different application of the volume operator, let us mention the proposal in \cite{Hartle:2007gi}. In that paper, they propose a resolution to the fact that the sphere produces a probability pressure towards smaller sizes via a term of the form 
$\exp\!\left( -d Q^2 \zeta_0 \right) $ \nref{Wliouv} \nref{Qsph}.
Their proposal was that probabilities conditioned on our observations should receive an additional volume weighting. The idea is that if the probability for an observer such as us to arise in any given Hubble-sized region is very small, then a universe with a larger reheating volume contains more opportunities for such an observer to arise. In this regime, the corresponding weighting is proportional to the total volume. To compare its dependence on the homogeneous mode with the sphere pressure, we write $\zeta = \zeta_0 + \widetilde{\zeta}$ and hold the nonzero modes $\widetilde{\zeta}$ fixed. The total probability then has the zero-mode dependence
\be 
 \int d^dx \sqrt{\hat \gamma } \exp \!\left(d b Q \zeta \right) \exp\!\left( -d Q^2 \zeta_0 \right) \propto  \exp \!\left(d b Q \zeta_0 -d Q^2 \zeta_0 \right) 
 \ee   
 where $\hat \gamma_{ij}$ is the metric of a unit sphere. 
 As such, the question is whether the positive volume weighting can balance the negative sphere pressure. However, after the renormalization \nref{LiouvRen}, the sphere pressure always dominates while $b$ is real, since $b < Q = b + 1/b$. Equivalently, the net zero-mode dependence is proportional to $\exp\!\left[-dQ(Q-b)\zeta_0\right] = \exp\!\left[-dQ\zeta_0/b\right]$, so the volume weighting does not reverse the preference for small reheating surfaces.
 One could still ask whether an analogous volume-weighting mechanism might operate in the regime of eternal inflation, where arbitrarily large reheating volumes become important. However, this is precisely the regime in which the renormalized volume operator description \nref{LiouvRen} breaks down, with $b$ becoming complex. It is therefore not clear how to define the corresponding volume-weighted probability in this regime, or whether the competition between the no-boundary weight and observer conditioning can be analyzed by the same argument.
 
 \subsection*{Acknowledgments}

We thank A.~Chang, E.~Gwynne, K.~Jensen, D.~Kolchmeyer, E.~Perlmutter, N.~Seiberg and E.~Witten for discussions.

\indent  

The work of JC is supported by the Simons Collaboration on Celestial Holography, as well as a Fellowship from the Alfred P.~Sloan Foundation.
The work of JM was supported in part by U.S. Department of Energy grant DE-SC0009988, the Carl Feinberg chair, and the Leinweber Forum at the IAS.

\appendix

\section{Checking consequences of the time reparametrization symmetry}
\la{CheckTime}

In this appendix, we derive \nref{ZetaWder}, which is a consequence of the bulk time reparametrization symmetry. 

\begin{figure}[t!]
    \centering
    \includegraphics[width=0.75\linewidth]{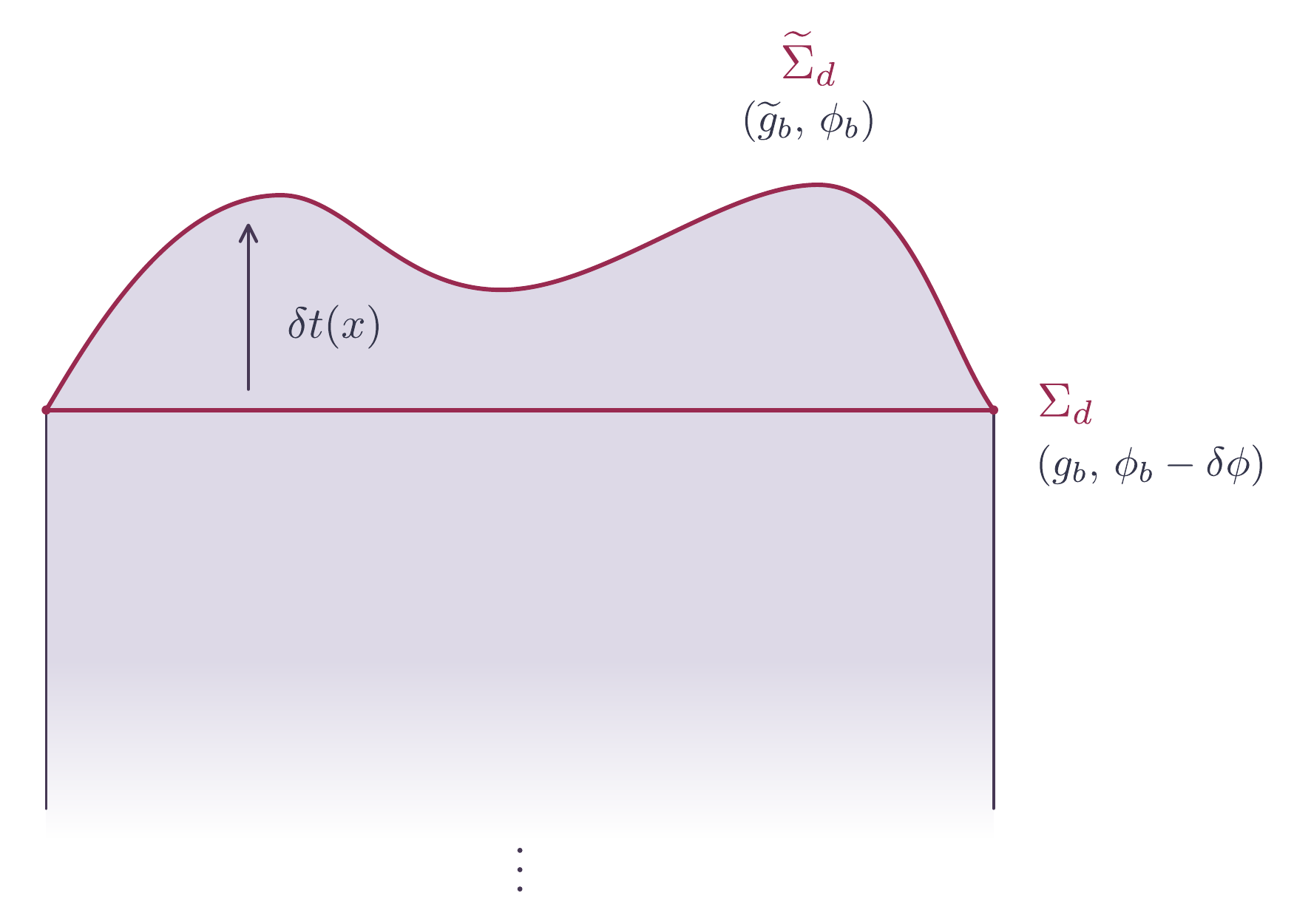}
    \caption{Example of the two surfaces $\Sigma_{d}$ and $\tilde{\Sigma}_{d}$ related by a local time evolution $\delta t(x)$ that we described in the text. Below $\Sigma_{d}$ is the rest of the no-boundary geometry.}
    \label{evolfig}
\end{figure}

We consider two set of boundary conditions related by a time reparametrization, see Figure \ref{evolfig}. 
One starts with a no-boundary manifold with reheating surface $\Sigma$ that has metric $g_{b}$. Then, one evolves the surface $\Sigma$ along its normal in \nref{metfg} by a local proper time $\delta t$, to a later surface $\tilde{\Sigma}$ with metric $\tilde{g}_{b}$, while changing the value of $\phi$ at $\Sigma$ so that $\phi=\phi_{b}$ at $\tilde{\Sigma}$. The net effect is that we obtain a new no-boundary geometry, with reheating surface $\tilde{\Sigma}$ (see Figure \nref{evolfig}). 

Its on-shell action $I_{\phi,+}[\tilde{g}_{b},\phi_{b}]$ can be computed as the sum of the action in the two regions of Figure \nref{evolfig}, namely the one below $\Sigma_{d}$ and the segment between $\tilde{\Sigma}_{d}$ and $\Sigma_{d}$. The contribution of the region below $\Sigma_{d}$ is the on-shell action of the no-boundary geometry with boundary $\Sigma_{d}$ but slightly displaced $\phi$, given by $I_{\phi,+}[g_{b},\phi_{b}-\delta \phi]$. The action of the segment follows from \nref{Iphi+} integrated between $\tilde{\Sigma}$ and $\Sigma$. The metrics $\tilde{g}_{b}$ and $g_{b}$ differ by $2K_{ij,+}\delta t$, where $K_{ij,+}$ is the extrinsic curvature of $\Sigma_{d}$ in the ket. In $\Sigma_{d}$ one has that $\delta \phi=\partial_{n}\phi_{+} \delta t$, with $\partial_{n}\phi_{+}$ the normal derivative of $\phi$ at $\Sigma_{d}$ computed in the ket. Comparing the two ways of computing the action, we obtain  
\begin{equation}
\label{halmid}
2K_{ij,+}\frac{\delta I_{\phi,+}}{\delta g_{ij}}\bigg|_{\phi=\phi_{r}}=\frac{1}{2}\sqrt{g_{b}}(\partial_{n}\phi_{+})^{2}-\partial_{n}\phi_{+} \frac{\delta I_{\phi,+}}{\delta \phi}\bigg|_{\phi=\phi_{r}}=-\frac{1}{2}\sqrt{g_{b}}(\partial_{n}\phi_{+})^{2}
\end{equation}
where on the right-hand side we used the simple identity $\frac{\delta I_{\phi,+}}{\delta \phi}=\sqrt{g_{b}}\partial_{n}\phi$ for a free scalar field $\phi$. The trick is then to write the equation below for $\CI_{\phi}$
\begin{equation}
\label{delIzeta}
\frac{\delta \CI_{\phi}}{\delta \zeta}\bigg|_{\phi=\phi_{r}}=\frac{i}{2}\sqrt{g_{b}}[(\partial_{n}\phi_{+})^{2}-(\partial_{n}\phi_{-})^{2}\big]+i\bigg[2(K_{ij}-g_{ij})_{+}\frac{\delta I_{\phi,+}}{\delta g_{ij}}\bigg|_{\phi=\phi_{r}}-2(K_{ij}-g_{ij})_{-}\frac{\delta I_{\phi,-}}{\delta g_{ij}}\bigg|_{\phi=\phi_{r}}\bigg]
\end{equation}
that can be integrated to find $\CI_{\phi}$ as a function of $\zeta$. The second term is subleading at small $\eta_{b}$. This is subtle because many terms in $\frac{\delta I_{\phi,\pm}}{\delta g_{ij}}$ diverge as we take $\eta_{b} \rightarrow 0$, but they are local functions of $g_{b}$, so they lead to terms common between the bra and the ket that cancel in \nref{delIzeta}. 

Then the first term in \nref{delIzeta} can be rewritten as \nref{ZetaWder} after using \nref{NormDerNFl} and \nref{Ndef}. To see why the second term is small, we can use some results of section \ref{eqwvfct} as follows. Note that 
\begin{equation}
\label{kijsimp}
(K_{ij}-g_{ij})|_{+}\frac{\delta I_{\phi,+}}{\delta g_{ij}}=\frac{1}{2}(-\eta_{b})\partial_{\eta}\gamma_{ij}(\eta_{b})|_{+}\frac{\delta I_{\phi,+}}{\delta \gamma_{ij}}
\end{equation}
where we wrote $K_{ij}$ explicitly in terms of $\gamma(\eta)$ in \nref{metfg} and wrote the derivative of the action with respect to the renormalized metric $\gamma_{b}$ instead, using the relation \nref{metbc}. We now should discuss how \nref{kijsimp} behaves as a power series in $\eta$. After using the equations of motion and integrating by parts, the on-shell action $I_{\phi,+}$ can be expressed in terms of $N$ in \nref{Ndef} as
\begin{equation}
\label{Ionshell}
I_{\phi,+}=-\frac{V'^{2}}{2d}\int_{\mathcal{M}}\frac{d\eta \, d^{d}x}{(-\eta)^{d+1}}\sqrt{\gamma(\eta)}\,N 
\end{equation}

The leading term in $I_{\phi,+}$ at small $\eta_{b}$ comes from the integral of $N$ near the boundary, since it behaves as approximately $\log(-\eta)-\log(-\eta_{b})$. The leading term in the action then goes as the volume of the reheating surface, which behaves as $(-\eta_{b})^{-d}\int_{\Sigma} d^{d}x\,\sqrt{\gamma_{b}}$. This term is a local function of the boundary metric, and so are the first subleading corrections to $I_{\phi,+}$ which still behave as a negative power of $\eta$. 

However, due to the filling dependencies of $N$ at order $\eta^{d}$, and due to the fact that $N$ in the interior grows with the cutoff as $-\log(-\eta_{b})$ from how much it rolled, the terms in $I_{\phi,+}$ that are $O(\log(-\eta_{b}))$ in the small $\eta$ expansion are the first ones that can be filling-dependent. With this in mind, one might wonder what the order of the leading filling dependent term in \nref{kijsimp} is. The leading filling dependent term in \nref{kijsimp} will come from the leading filling-dependent piece in either the derivative of $I_{\phi,+}$ or from the $\partial_{\eta}\gamma(\eta)$ term. The leading filling dependent piece of $(-\eta)\partial_{\eta}\gamma$, from \nref{gammadirexp}, is of order $\eta^{d}$, and is proportional to the coefficient $h_{d}$ in \nref{gammadirexp}. The leading term from the derivative of $I_{\phi,+}$ is of order $(-\eta)^{-d}$, so it looks like it could combine with this term to give a filling-dependent piece of $O(1)$. However, since the $(-\eta)^{-d}$ term comes from the volume, its derivative with respect to the metric is proportional to $(\gamma_{b})^{ij}$, and when contracted with $h_{d}$ it gives something local, due to the constraint \nref{trhd}. 

The leading filling-dependent piece of $\frac{\delta I}{\delta \gamma}$ is of order $O(\log(-\eta_{b}))$ as we discussed, but the leading term of $(-\eta)\partial_{\eta}\gamma$ is $O(\eta^{2})$, so this term does not give a finite contribution at small $\eta_{b}$. In conclusion, the leading filling-dependent contributions from both terms in \nref{kijsimp} are less than $O(1)$ in the small $\eta$ expansion, and thus do not survive the small $\eta$ limit. This allows us therefore to conclude that
\begin{equation}
\label{kijsmall}
(K_{ij}-g_{ij})|_{+}\frac{\delta I_{\phi,+}}{\delta g_{ij}}-(K_{ij}-g_{ij})|_{-}\frac{\delta I_{\phi,-}}{\delta g_{ij}}=o(1)
\end{equation}
as we wanted to argue.

\section{The sphere zero mode from the flat space Green function}
\label{SphereCharge}

In section \ref{sphex} we computed ${\cal Q}_{S^d}$ directly from the de Sitter filling and found  ${\cal Q}_{S^d} = \Gamma(d)$, which implies the linear term ${\cal I}_{\ell} = d Q^2\zeta_0 + \cdots$ in \nref{Qsph}. Here we give another way of obtaining this normalization. The argument makes clear that the coefficient of the sphere zero mode and the coefficient of the logarithm in the short distance two-point function are fixed by the same normalization of ${\cal P}$.

Write the metric of the unit round sphere in stereographic coordinates as
\begin{align}
\label{StereographicCharge}
d\hat s^2_{S^d} = e^{2\sigma(x)}d x^i d x^i ~,~~~~~~ e^{\sigma(x)}={ 2 \over 1+|x|^2} \,.
\end{align}
At large $|x|$,
\begin{align}
\label{SigmaAsymCharge}
\sigma(x)=-2\log |x|+\log 2+O(|x|^{-2}) \,.
\end{align}

For the de Sitter filling considered in section \ref{sphex}, let ${\cal P}_{\delta}$ denote the operator ${\cal P}$ after stereographically mapping the sphere to the flat metric $\delta_{ij}$. In this case, conformal covariance together with \nref{Psph} gives
\begin{align}
\label{PflatExactCharge}
{\cal P}_{\delta} = (-\nabla^2)^{d/2} \,,
\end{align}
where for odd $d$ the right hand side is understood as the fractional Laplacian. We will use that ${\cal P}_{\delta}$ is self-adjoint, homogeneous of order $d$, and annihilates constants:
\begin{align}
\label{PoneCharge}
{\cal P}_{\delta}1=0 \,.
\end{align}

We first fix the normalization of the logarithmic fundamental solution from the two-point function. In flat space, the quadratic part of \nref{Wliouv} is
\begin{align}
\label{FlatActionCharge}
{\cal I}_{\ell}^{(2)}={ dQ^2 \over 2\Gamma(d){\rm Vol}(S^d)}\int d^d x\,\zeta {\cal P}_{\delta}\zeta \,.
\end{align}
Therefore, defining $G(x)=\langle \zeta(x)\zeta(0)\rangle$, we have
\begin{align}
\label{GreenEqCharge}
{ dQ^2 \over \Gamma(d){\rm Vol}(S^d)}{\cal P}_{\delta}\,G(x)=\delta^{(d)}(x) \,.
\end{align}
Since ${\cal P}_{\delta}=(-\nabla^2)^{d/2}$ is scale invariant, its flat space Green function is logarithmic, up to an additive constant and an infrared prescription. Matching its normalization to the short distance result \nref{zetatwopt}, we find
\begin{align}
\label{LogCharge}
{\cal P}_{\delta}\left(-2\log |x|\right)=\Gamma(d){\rm Vol}(S^d)\,\delta^{(d)}(x) \,.
\end{align}

We now use the Weyl transformation law \nref{QPRel}. Since the flat metric has ${\cal Q}_{\delta}=0$, \nref{QPRel} and \nref{StereographicCharge} give
\begin{align}
\label{QPsigmaCharge}
\sqrt{\hat \gamma_{S^d}}\,{\cal Q}_{S^d}={\cal P}_{\delta}\sigma \,,
\end{align}
and so the total ${\cal Q}$-curvature of the sphere can be computed from the total ${\cal P}_{\delta}$ charge of the stereographic Weyl factor.

For even $d$, the total ${\cal P}_{\delta}$ charge of the stereographic Weyl factor can be evaluated by ordinary integration by parts. It is useful, however, to give a version of the argument that also applies when $d$ is odd and ${\cal P}_{\delta}$ is non-local. Let $\rho$ be a smooth, compactly supported function that is equal to one near the origin, and define $\rho_R(x) = \rho(x/R)$. We define the total charge by
\begin{align}
\label{TotalChargeDef}
I[\sigma]\equiv\lim_{R\to\infty}\left\langle {\cal P}_{\delta}\sigma,\rho_R\right\rangle .
\end{align}
Here $\langle \cdot\,,\,\cdot\rangle$ denotes the usual bilinear pairing.  Using self-adjointness,
\begin{align}
\left\langle {\cal P}_{\delta}\sigma,\rho_R\right\rangle=\left\langle \sigma,{\cal P}_{\delta}\rho_R\right\rangle \,.
\end{align}
Since ${\cal P}_{\delta}$ is homogeneous of order $d$,
\begin{align}
{\cal P}_{\delta}\rho_R(x)=R^{-d}({\cal P}_{\delta}\rho)(x/R) \,.
\end{align}
Changing variables to $x=Ry$ gives
\begin{align}
\label{RescaledCharge}
\left\langle \sigma,{\cal P}_{\delta}\rho_R\right\rangle=\int_{\mathbb{R}^d}d^d y\,\sigma(Ry)({\cal P}_{\delta}\rho)(y) \,.
\end{align}
For the stereographic Weyl factor,
\begin{align}
\label{SigmaRescaledCharge}
\sigma(Ry)=\log 2-\log(1+R^2|y|^2)=-2\log R+\log 2-\log\!\left(|y|^2+R^{-2}\right) \,.
\end{align}
The terms independent of $y$ do not contribute because
\begin{align}
\int_{\mathbb{R}^d}d^d y\,{\cal P}_{\delta}\rho=\left\langle 1,{\cal P}_{\delta}\rho\right\rangle=\left\langle {\cal P}_{\delta}1,\rho\right\rangle=0 \,.
\end{align}
Taking $R\to\infty$ in \nref{RescaledCharge}, we therefore obtain, in the distributional sense,
\begin{align}
I[\sigma]=\int_{\mathbb{R}^d}d^d y\,(-2\log |y|)({\cal P}_{\delta}\rho)(y)=\left\langle {\cal P}_{\delta}(-2\log |y|),\rho\right\rangle \,.
\end{align}
Using \nref{LogCharge} and $\rho(0)=1$, we find
\begin{align}
\label{TotalChargeResult}
I[\sigma]=\Gamma(d){\rm Vol}(S^d) \,.
\end{align}
Combining this with \nref{QPsigmaCharge} gives
\begin{align}
\label{SphereQIntegralCharge}
\int_{S^d}d^d x\sqrt{\hat \gamma}\,{\cal Q}_{S^d}=\Gamma(d){\rm Vol}(S^d) \,.
\end{align}
Since ${\cal Q}_{S^d}$ is constant by rotational invariance, this reproduces ${\cal Q}_{S^d}=\Gamma(d)$, in agreement with the direct bulk calculation \nref{Qsph}.

For a constant configuration $\zeta=\zeta_0$, the quadratic term vanishes because ${\cal P}1=0$. Using \nref{SphereQIntegralCharge} in \nref{Wliouv}, we obtain
\begin{align}
\label{SphereZeroFromGreen}
{\cal I}_{\ell}[\zeta_0,S^d]={ dQ^2 \over 2\Gamma(d){\rm Vol}(S^d)}\left[2\zeta_0\int_{S^d}d^d x\sqrt{\hat \gamma}\,{\cal Q}_{S^d}\right]=dQ^2\zeta_0 \,.
\end{align}
Together with \nref{zetatwopt}, this gives
\begin{align}
\label{SphereTwoptIdentity}
\left({ \partial {\cal I}_{\ell} \over \partial\zeta_0}\right)\lim_{|x|\to0}\left[{\langle\zeta(x)\zeta(0)\rangle \over \log { 1 \over |x|}}\right]=(dQ^2)\left({ 2 \over dQ^2}\right)=2 \,.
\end{align}
Equation \nref{SphereTwoptIdentity} makes explicit the relationship between the coefficient of the sphere zero mode and the normalization of the short distance logarithmic fluctuations.

For even $d=2m$, the same argument can be expressed as an ordinary surface integral because ${\cal P}_{\delta}=(-\nabla^2)^m$ is local. For a ball $B_R$,
\begin{align}
\int_{B_R}d^d x\,{\cal P}_{\delta}f=(-1)^m\int_{\partial B_R}dS\,\partial_r(\nabla^2)^{m-1}f \,.
\end{align}
The cutoff argument above is the analogous statement that also applies for odd $d$, where ${\cal P}_{\delta}$ is a non-local fractional operator.

We therefore see that the sphere zero mode and the short distance logarithmic fluctuations are not independently normalized. The same flat space Green function normalization, together with conformal covariance, fixes both the coefficient $dQ^2$ of the sphere pressure and the coefficient $2/(dQ^2)$ of the logarithmic two-point function.

\section{No-boundary density matrix for the inflaton}
\label{denmat}

In section \ref{nbdinfl} we discussed the wavefunctional weight $\cal{I}_{\phi}$ of $\zeta$ in the reheating surface for a given bra-ket pair in the wavefunctional. However, from the perspective of a late-time observer such as us, the more natural object to consider is instead a density matrix $\rho[\zeta_{+},\zeta_{-}]$ of our observable patch $\Sigma_{\text{in}}$ of the reheating surface \cite{Ivo:2024ill}, with $\zeta_{\pm}$ boundary configurations in the ket and bra, respectively. Here we will restrict ourselves to the diagonal components of the density matrix, with $\zeta_{+}=\zeta_{-}$. The diagonal density matrix of a given region $\Sigma_{\text{in}}$ can be obtained by taking the wavefunctional squared $|\Psi|^{2}$ in \nref{psibraket} for $\Sigma_{d}=\Sigma_{\text{in}}\cup \Sigma_{\text{out}}$, and integrating out the degrees of freedom in $\Sigma_{\text{out}}$.  

A simple comment is that in the situations where $\mathcal{P}_{\gamma_{b}}$ is local, as those discussed in section \ref{prevdef}, the density matrix as a function of $\zeta$ inside of $\Sigma_{\text{in}}$ has the same form as \nref{WActL}, due to locality of the problem\footnote{A wrinkle is that the integrating-out procedure can produce terms that depend on data in the boundary of $\Sigma_{\rm in}$.}. However, the integrating-out procedure is non-trivial for most bra-ket fillings, where the kinetic operator $\mathcal{P}_{\gamma_{b}}$ in \nref{WActL} is generically non-local. This implies that integrating out the exterior for such saddles changes the final dependence of the density matrix on the $\zeta$ inside the region $\Sigma_{\text{in}}$. 

Another comment is that in the context of the density matrix, the region $\Sigma_{\text{in}}$ generically has a boundary $\mathcal{C}=\partial \Sigma_{\text{in}}$, which corresponds to the intersection of the light cone of the late time observer with the reheating surface (see Figure \nref{cosmogic}). One might therefore, as considered in \cite{Ivo:2024ill}, be interested in the behavior of the density matrix for a given area $A_{\mathcal{C}}$ of $\mathcal{C}$. In that case, it is convenient to decompose $\zeta$ as
\begin{equation}
\label{zetasplitA}
\zeta=\zeta_{f}+\zeta_{A}
\end{equation}
where $\zeta_{A}$ is a constant that controls the area of $\mathcal{C}$, and $\zeta_{f}$ describes fluctuations of $\zeta$ in $\Sigma_{\text{in}}$ that do not change the area. Interestingly, a similar type of decomposition also appears naturally for Liouville theory \cite{Chatterjee:2024phq} on the sphere, where one decomposes the Liouville field into a zero mode plus a free field component which has zero ``circle average'' at the equator. Here, this decomposition appears naturally from the perspective of a late-time observer. 

To make the connection with \cite{Chatterjee:2024phq} more manifest, we restrict ourselves to the sphere boundary with a de Sitter filling, and to radially symmetric configurations of $\zeta$. In practice this means we take the metric $\hat{\gamma}_{b}$ to be given by
\begin{equation}
d\hat{s}^{2}=d\theta^{2}+\sin^{2}\theta\,d\Omega_{d-1}^{2}
\end{equation}
and we consider configurations $\zeta_{f}(\theta)$ that depend on $\theta$ alone. Moreover, by using the symmetries of de Sitter, one can set the boundary $\mathcal{C}$ to be at $\theta=\frac{\pi}{2}$ \cite{Ivo:2024ill}. For such configurations, our split \nref{zetasplitA} is therefore equivalent to that of \cite{Chatterjee:2024phq}. Following \cite{Chatterjee:2024phq}, one is interested in expanding $\zeta_{f}=\tilde{\zeta}_{f}+\delta \zeta_{f}$ around a configuration $\tilde{\zeta}_{f}$ that vanishes at $\mathcal{C}$ and extremizes the action for a given $\zeta_{A}$, in such a way that the action for $\delta \zeta_{f}$ becomes purely quadratic.  Extremizing \nref{WActL}, one obtains that away from $\mathcal{C}$ such $\tilde{\zeta}_{f}$ satisfies
\begin{equation}
\label{Qfree}
\mathcal{Q}_{\hat{\gamma}_{b}}+\mathcal{P}_{\hat{\gamma}_{b}}\tilde{\zeta}_{f}=e^{d\tilde{\zeta}_{f}}\mathcal{Q}_{e^{2\tilde{\zeta}_{f}}\hat{\gamma}_{b}}=0~~\text{ away from $\partial \Sigma_{\text{in}}$}
\end{equation}

Therefore, the $\tilde{\zeta}_{f}$ corresponds to metrics with ``zero curvature'', as measured by $\mathcal{Q}_{\gamma_{b}}$. In the context of sphere boundaries with a de Sitter filling in even $d$, this is equivalent to metrics with zero Branson $Q$-curvature. In $d=2$, for example, the Branson $Q$-curvature is proportional to the Ricci scalar, so \nref{Qfree} is the condition that the metric is Ricci flat away from the boundary. In that case, $\tilde{\zeta}_{f}$ is such that $e^{2\tilde{\zeta}_{f}}d\hat{s}^{2}$ is the metric of two flat disks glued at their boundary, as it was also discussed in \cite{Chatterjee:2024phq}. In higher dimensions, however, the solutions to \nref{Qfree} are more complicated. The issue is that solutions that are flat in $\Sigma_{\text{in}}$ and $\Sigma_{\text{out}}$ would be too discontinuous at their junction.

To illustrate this, we now solve the variational problem more explicitly.  The variational equation \nref{Qfree} for $\tilde{\zeta}_{f}$ is allowed to have a delta function at $\theta=\frac{\pi}{2}$, and for radially symmetric configurations we have that on the sphere
\begin{equation}
\label{eqtildez}
\mathcal{P}_{\hat{\gamma}_{b}}\tilde{\zeta}_{f}+\Gamma(d)=\frac{\Gamma(d)\text{Vol}(S^{d})}{\text{Vol}(S^{d-1})}\,\delta\bigg(\theta-\frac{\pi}{2}\bigg)
\end{equation}
where the constant on the right-hand side can be obtained by integrating both sides along the sphere and using that $\mathcal{P}_{\hat{\gamma}_{b}}1=0$. We can write the solution to \nref{eqtildez} in terms of spherical harmonics $Y_{l}(\theta)$ that depend just on $\theta$ as
\begin{equation}
\tilde{\zeta}_{f}=\Gamma(d)\text{Vol}(S^{d})\sum_{l>1} \frac{\Gamma(l)Y_{l}\big(\frac{\pi}{2}\big)}{\Gamma(l+d)}\big(Y_{l}(\theta)-Y_{l}(\frac{\pi}{2})\big)
\end{equation}

By evaluating this expression for various dimensions, one finds
\begin{equation}
\widetilde{\zeta}_{f}(\theta)=
\begin{cases}
-\log(1+|\cos\theta|),
& d=2, \\[10pt]
\log 2-
\dfrac{
(1+\sin\theta)\log(1+\sin\theta)
-(1-\sin\theta)\log(1-\sin\theta)
}{
2\sin\theta
},
& d=3, \\[12pt]
-\log(1+|\cos\theta|)
+\dfrac{|\cos\theta|}{1+|\cos\theta|},
& d=4.
\end{cases}
\label{zetafex}
\end{equation}
where the $d=2$ solution for $\tilde{\zeta}_{f}$ is precisely the one necessary for the metric to be that of a flat disk. The issue of using it in higher dimensions is that its discontinuities in \nref{eqtildez} would be more severe than delta functions. Indeed, for $\theta=\frac{\pi}{2}\pm \epsilon$ the flat disk expression behaves as $\zeta_{f}\approx -|\epsilon|$, so its first derivative is discontinuous at $\theta=\frac{\pi}{2}$. The actual higher-dimensional solutions in \nref{zetafex} have continuous first derivatives, with discontinuities or divergences appearing only at derivatives of order $d-1$, which is necessary for \nref{eqtildez} to hold.

Another comment is that in even $d=2k$ there is a more elegant way of solving \nref{eqtildez}. The trick is to take $\tilde{\zeta}_{f}=-\log(1+\cos \theta)+\hat{\zeta}_{f}$ and $r=\tan \frac{\theta}{2}$, so that $r$ is between $0$ and $1$ for $\theta$ between $0$ and $\frac{\pi}{2}$. In terms of these variables, the metric $e^{2\tilde{\zeta}_{f}}d\hat{s}^{2}$ is
\begin{equation}
ds^{2}=e^{2\tilde{\zeta}_{f}}d\hat{s}^{2}=e^{2\hat{\zeta}_{f}}(dr^{2}+r^{2}d\Omega_{d-1}^{2})
\end{equation}
so equation \nref{eqtildez} becomes
\begin{equation}
(\nabla^2)^k \hat \zeta_{f} = \bigg(\frac{1}{r^{2k-1}}\frac{d}{d r}\bigg(r^{2k-1}\frac{d}{dr}\bigg)\bigg)^{k}\hat{\zeta}_{f}(r)=(-1)^{k}2^{2k-1}[(k-1)!]^{2}\delta(r-1)
\end{equation}
And the solution for $\hat{\zeta}_{f}$ at $r>1$ is defined by $\hat{\zeta}_{f}(r)=\hat{\zeta}_{f}(r^{-1})-2\log r$.

\section{No-boundary wavefunction for a massive scalar field}
 \la{MassFie}

Here we discuss the action $\CI_{\phi}$ when we take the potential of $\phi$ to be of the form
\begin{equation}
\label{potm}
V(\phi)=V_{0}+\frac{1}{2}m^{2}\phi^{2}~,~~~~{\rm with }
~~~ V_0 \equiv \frac{d(d-1)}{2}M_{\rm pl}^{d-1} 
\end{equation}
and we take $M_{\rm pl} \rightarrow \infty$ with $H$ and $m$ fixed. For similar reasons as discussed in section \ref{gennb}, the action for the scale factor $\zeta$ of the reheating surface comes purely from $\CI_{\phi}$. If the metric is asymptotically of the form \nref{metfg}, then, as is usually discussed in the AdS context for a massive scalar field \cite{Breitenlohner:1982jf, Mezincescu:1984ev, Witten:1998qj, Gubser:1998bc, Graham:2003GZ}, the asymptotic behavior of $\phi$ is given by
\begin{equation}
\label{mexp}
\phi=\chi_{\rm s}(x)(-\eta)^{\Delta}[1+ \cdots ]+\chi_{\rm f}(x)(-\eta)^{d-\Delta}[1+ \cdots ]
\end{equation}
where the dotted terms are locally determined by the respective $\chi$, and
\begin{equation}
\Delta=\frac{d}{2}- \sqrt{\bigg(\frac{d}{2}\bigg)^{2}-m^{2}}
\end{equation}
so that $\chi_{\rm f}$ and $\chi_{\rm s}$ are the ``fast'' and ``slow'' decaying solutions, respectively. Note also that $m^{2}=\Delta(d-\Delta)$, and we assume $0<m^{2}<\frac{d^{2}}{4}$. The slow solution controls the decay of $\phi$ at small enough $\eta$, and therefore the asymptotic value of $\phi$ fixes $\chi_{\rm s}$ locally. Namely, if $\phi$ is fixed to a value $\phi(\Sigma)$ in a surface at $\eta=\eta_{b}$ we have that
\begin{equation}
\label{chimfix}
\chi_{\rm s}\approx (-\eta_{b})^{-\Delta}\phi(\Sigma)
\end{equation}
while $\chi_{\rm f}$ is fixed by requiring regularity of \nref{mexp} inside the no-boundary geometry, being therefore filling dependent. 

As in section \ref{nbdinfl}, we will be interested in studying the wavefunctional at a surface of fixed $\phi=\phi_{r}$, which we take to respect
\begin{equation}
\partial_{\phi}V|_{\phi=\phi_{r}}=m^{2}\phi_{r}=V_{r}' \rightarrow \phi_{r}=\frac{V_{r}'}{\Delta(d-\Delta)}
\end{equation}
with $V_{r}'$ a fixed constant, so that near $\phi=\phi_{r}$ the problem is similar to the linear one we discussed in \nref{nbdinfl}. On this surface, at a small value of $\eta=\eta_{b}$, we take the renormalized metric to be of the form $\gamma_{b}=e^{2\zeta}\hat{\gamma}_{b}$ and we want to compute the action of $\zeta$. 

The simplest way to do this is by using the same trick as in \nref{ZetaWder}, where we relate the variation of the action with respect to $\phi$ to the variation with respect to $\zeta$, using the asymptotic symmetry of the problem. However, here the symmetry of the wavefunctional is different than \nref{weyl1}. The asymptotic symmetry analog to \nref{weyl1} follows from \nref{mexp} to be
\begin{equation}
\label{weylm}
\zeta \rightarrow \zeta +\omega~,~~~\phi \rightarrow \phi \,e^{-\Delta\omega}
\end{equation}
which implies that
\begin{equation}
\label{ZetaWderm}
\frac{\delta \CI_{\phi}}{\delta \zeta}\bigg|_{\phi=\phi_{r}}=\Delta \phi_{r}\frac{\delta \CI_{\phi}}{\delta \phi}=-i\Delta \phi_{r}(-\eta_{b})^{-d}\sqrt{\gamma_{b}}(\partial_{n}\phi_{+}-\partial_{n}\phi_{-})
\end{equation}
where we use the variation of the on-shell action with respect to $\phi$ for a free scalar, similarly to \nref{PhiWder}. We parameterize the bra-ket discontinuity of the normal derivative as
\begin{equation}
\label{discfm}
\partial_{n}\phi_{+}-\partial_{n}\phi_{-}=2iC_{\Delta}(-\eta_{b})^{d-\Delta}\mathcal{P}_{d-2\Delta,\gamma_{b}}\chi_{\rm s}+ \cdots ~,~~ C_{\Delta}=\frac{\pi}{2^{d-2\Delta-1}\Gamma^{2}\big(\frac{d}{2}-\Delta\big)}
\end{equation}
where the $\mathcal{P}_{d-2\Delta,\gamma_{b}}$ is a linear operator that acts on $\chi_{\rm s}\approx (-\eta_{b})^{-\Delta}\phi(\Sigma)$ in \nref{mexp} to produce, up to prefactors, the difference in the fast mode response $\chi_{\rm f}$ between the ket and bra. The power of $(-\eta_{b})$ in \nref{discfm} follows from the power of $(-\eta)$ in front of $\chi_{\rm f}$ in \nref{mexp}. Taking into account how $\chi_{\rm s}$ is related to $\phi_{b}$ in \nref{chimfix}, one sees that \nref{discfm} scales as $(-\eta_{b})^{d-2\Delta}$ for a given $\phi_{b}$. By direct comparison with \nref{fdisc} at $m^{2}=\Delta=0$, we see that $\mathcal{P}_{d,\gamma_{b}}$  is equal to what we previously called $\mathcal{P}_{\gamma_{b}}$. So, we can think of $\mathcal{P}_{d-2\Delta,\gamma_{b}}$ as a generalization of the response $\mathcal{P}_{\gamma_{b}}$ for non-zero mass. 

To integrate \nref{ZetaWderm}, we should discuss how $\mathcal{P}_{d-2\Delta,\gamma_{b}}$ transforms under Weyl transformation of $\gamma_{b}$. A quick way to argue for this non-rigorously using ideas of section \ref{folweyl} is by noting that taking $\eta_{b} \rightarrow \eta_{b}e^{\omega}$ and changing the renormalized metric as $e^{2\omega}\gamma_{b}$ should lead to a problem equivalent to the original one. Keeping $\phi_{b}$ fixed changes $\chi_{\rm s}$ from \nref{chimfix} by a factor of $e^{-\Delta\omega}$. Since $\mathcal{P}_{d-2\Delta,\gamma}$ maps $\chi_{\rm s}$ to the difference of $\chi_{\rm f}$, comparing \nref{discfm} between the problem with $\gamma_{b}$ and the problem with $e^{2\omega}\gamma_{b}$ we have that
\begin{equation}
\label{weylPm}
\mathcal{P}_{d-2\Delta,e^{2\omega}\gamma_{b}}(e^{-\Delta \omega}f)=e^{-(d-\Delta)\omega}\mathcal{P}_{d-2\Delta,\gamma_{b}}f
\end{equation}

Note that for $\Delta=0$, \nref{weylPm} reduces to the Weyl transformation of $\mathcal{P}_{\gamma_{b}}$ in \nref{Pweyl}. One can also show that $\mathcal{P}_{d-2\Delta,\gamma_{b}}$ satisfies
\begin{equation}
\label{Pmsym}
\int_{\Sigma} d^{d}x\sqrt{\gamma_{b}}\,f \mathcal{P}_{d-2\Delta,\gamma_{b}}h=\int_{\Sigma} d^{d}x\sqrt{\gamma_{b}}\,h\mathcal{P}_{d-2\Delta,\gamma_{b}}f
\end{equation}

Integrating \nref{ZetaWderm} starting from $\zeta=0$ at fixed $\phi_{b}=\phi_{r}$, and using \nref{weylPm} and \nref{Pmsym}, we derive that
\begin{equation}
\label{actm}
\CI_{\phi}[e^{2\zeta}\hat{\gamma}_{b}]=\CI_{\phi}[\hat{\gamma}_{b}]+C_{\Delta}\bigg(\frac{V_{r}'}{\Delta(d-\Delta)}\bigg)^{2}(-\eta_{b})^{-2\Delta}\int_{\Sigma} d^{d}x\,\sqrt{\hat{\gamma}_{b}}\big[e^{\Delta \zeta}\mathcal{P}_{d-2\Delta,\hat{\gamma}_{b}}e^{\Delta\zeta}-\mathcal{P}_{d-2\Delta,\hat{\gamma}_{b}}1\big]
\end{equation}

An issue with the action \nref{actm} is that it does not have a finite limit as we take $\eta_{b} \rightarrow 0$ with the other parameters fixed. This is clear mathematically from the fact that the action is a function only of the actual physical Weyl factors of the metric, which are of the form $\zeta-\log(-\eta_{b})$, and $\zeta$ appears exponentially in \nref{actm}. Physically, this has to do with the fact that since the potential \nref{potm} is quadratic, taking $\eta_{b}$ small probes large $\phi$ regions in the past where the slope is big, so the associated fluctuations are small. However, the action \nref{actm} acts at least as an interesting curiosity of an action that is conformally invariant but different from \nref{WActL}. To check for conformal invariance, note that under conformal transformation \nref{SpConf}, we have that
\begin{equation}
f_{\zeta}=e^{\Delta \zeta} \rightarrow \delta f_{\zeta}=\xi^{i}\hat{\nabla}_{i}f_{\zeta}+\frac{\Delta}{d}\hat{\nabla}_{i}\xi^{i}\,f_{\zeta}
\end{equation}

Moreover, using \nref{weylPm} and diffeomorphism covariance, we can conclude that
\begin{equation}
\delta (\mathcal{P}_{d-2\Delta,\hat{\gamma}_{b}}f_{\zeta})=\xi^{i}\hat{\nabla}_{i}(\mathcal{P}_{d-2\Delta,\hat{\gamma}_{b}}f_{\zeta})+\frac{(d-\Delta)}{d}\hat{\nabla}_{i}\xi^{i}\, \mathcal{P}_{d-2\Delta,\hat{\gamma}_{b}}f_{\zeta}
\end{equation}
therefore the action varies as
\begin{equation}
\delta \CI_{\phi}=C_{\Delta}\bigg(\frac{V_{r}'}{\Delta(d-\Delta)}\bigg)^{2}(-\eta_{b})^{-2\Delta}\int_{\Sigma}d^{d}x\sqrt{\hat{\gamma}_{b}}[\xi^{i}\hat{\nabla}_{i}(f_{\zeta}\mathcal{P}_{d-2\Delta,\hat{\gamma}_{b}}f_{\zeta})+\hat{\nabla}_{i}\xi^{i}\,f_{\zeta}\mathcal{P}_{d-2\Delta,\hat{\gamma}_{b}}f_{\zeta}]=0
\end{equation}
where we used that the integrand is a total derivative.

 We remark that by taking a $\Delta \rightarrow 0$ limit at fixed $V_{r}'$ and $\eta_{b}$, one can show that \nref{actm} reduces to the linear-roll action in \nref{WActL}. To take this limit, one has to note that $\mathcal{P}_{d-2\Delta,\gamma}|_{\Delta=0}=\mathcal{P}_{\gamma}$, and $\partial_{\Delta}\mathcal{P}_{d-2\Delta,\gamma}1|_{\Delta=0}=\mathcal{Q}_{\gamma}$. The latter follows by a simple extension of the argument in section 4 of \cite{Fefferman:2002QCurvaturePoincare} to our context. 

\section{Review of one-loop determinants on $H^3$}
\label{app:H3det}

Here we briefly review the $H^3$ determinant calculation used in section~\ref{loops}, following~\cite{Giombi:2008vd,Cotler:2018zff}. We set the $H^3$ radius to one.

For a massless scalar, the coincident heat kernel is
\begin{equation}
K_{\phi}(s;x,x) = \frac{e^{-s}}{(4\pi s)^{3/2}}\,.
\end{equation}
Since $H^3$ is homogeneous, the coincident heat kernel is position independent, so
\begin{equation}
\text{Tr} \, e^{-s\mathcal{O}} = \int_{H^3_{\rm reg}} d^3 x\,\sqrt{g}\,\text{tr}\,K_{\mathcal{O}}(s;x,x) = V_{\rm reg}\,\text{tr}\, K_{\mathcal{O}}(s;x,x)\,.
\end{equation}
Therefore, using
\begin{equation}
\log A_\phi = \frac{1}{2}\int_0^\infty \frac{ds}{s}\,\operatorname{Tr}\!\left(e^{-s\mathcal{O}_\phi}\right)\,,
\end{equation}
we find
\begin{equation}
\log A_{\phi,H^3}^{\rm reg} = \frac{V_{\rm reg}}{2(4\pi)^{3/2}} \int_0^\infty ds\,s^{-5/2} e^{-s} = \frac{1}{12\pi} \,V_{\rm reg}\,,
\label{H3scalar}
\end{equation}
where the final equality is understood by analytic continuation.

For gravity, including the bulk Faddeev-Popov ghosts, the combined heat kernel calculation gives~\cite{Giombi:2008vd, Cotler:2018zff}
\begin{equation}
\log A_{{\rm grav},H^3}^{\rm reg} = V_{\rm reg} \int_0^\infty \frac{ds}{s}\frac{1}{(4\pi s)^{3/2}} \left[e^{-s} (1 + 8s) - e^{-4s} (1 + 2s)\right] = -\frac{13}{6\pi} \,V_{\rm reg}\,,
\label{H3gravity}
\end{equation}
up to Weyl-independent terms and phases. These are the quantities used in section~\ref{loops}.

The coefficient $13$ is the familiar one-loop contribution to the effective central charge in three-dimensional gravity~\cite{Cotler:2018zff}.

\bibliographystyle{apsrev4-1long}
\bibliography{main.bib}

\end{document}